
\input amstex
\documentstyle{amsppt}
\document
\NoBlackBoxes
\NoRunningHeads
\TagsAsMath
\TagsOnRight
\font\ninebf=cmbx9
\font\nineit=cmti9
\font\special=cmr6
\topmatter
\title
Form Factors, deformed Knizhnik-Zamolodchikov equations\\
and finite-gap integration.
\endtitle
\author
Fedor A. Smirnov\\
{\special Isaac Newton Institute, Cambridge, CB3 0EH, U.K.}\\
{\special and}\\
{\special Steklov Mathematical Institute,
  St. Petersburg 191011,  RUSSIA}
\endauthor
\abstract
We study the limit of asymptotically free massive integrable
models in which the algebra of nonlocal charges turns into
affine algebra. The form factors of fields in that limit
are described by KZ equations on level 0. We show the limit
to be connected with finite-gap integration of classical
integrable equations.
\endabstract
\endtopmatter
{\bf 1.Motivations.}

\define\Tau{\frak{V}}
\define\DET{\text{det}}
\define\EXP{\text{exp}}
\define\la#1{\lambda_{#1}}
\define\laaa#1{\lambda_1 ,\cdots , \lambda_{#1}}
\define\gaaa#1{\gamma_1,\cdots,\gamma_{#1}}
\define\teet#1#2{\theta [\eta _{#1}] (#2)}
\define\tede#1{\theta [\delta](#1)}
\define\taaa #1{\tau _1,\cdots, \tau _{#1}}

The nonlocal symmetries of integrable models of
quantum field theory in two dimensions were first studied
several years ago [1]. The reason for that being in
attempts to understand the quantization of asymptotically
free models. Being almost forgotten for some time the
nonlocal symmetries returned to the field rather indirectly,
namely, through the finite-dimensional quantum group
symmetries of CFT. Now it is understood that the
integrability of massive models is closely
connected with possessing infinite-dimensional algebra
of nonlocal symmetries [2-6]. The local integrals constitute
a center of it. The algebra of nonlocal
symmetries is always a Hopf algebra, particles
transform under its finite-dimensional representations
while quasilocal fields constitute infinite-dimensional multiplets
with highest vectors corresponding to the local
fields. The S-matrix is nothing but
universal R-matrix specified onto finite-dimensional representations
while the braiding of the multiplets of quasilocal fields is
described by the universal R-matrix specified onto the
tensor product of two Verma modules.
Moreover, it was shown [6] that the form factors which
put together particles and fields can be considered as
solutions of deformed Knizhnik-Zamolodchikov equations [7]
which is quite natural since these equations use to
relate finite-dimensional representations of deformed
loop algebras with infinite-dimensional highest weight
representations.

A question which should be asked is the following.
Suppose we have an integrable model with
symmetry under certain deformation of the
loop algebra. Then what is the meaning of the
``classical limit'' which moves the deformed
algebra into undeformed one. The answer to this
question is not trivial as we shall see.
For many models this limit does not
look to make much sense.
The point is that the limit is related not to
the rescaling of space coordinates (or momenta) as conformal
limit is, but rather to rescaling of rapidities of particles
which are logarithms of momenta. In a sense the limit in question is even
opposite to conformal limit in which we throw away all the
logarithms preserving powers, here we preserve logarithms,
but throw away powers considering the asymptotic expansions.
However, the limit seems to
be reasonable for asymptotically free models.
So, we return to the same idea as years ago:
to use nonlocal symmetries in asymptotically free theories,
certainly we return to this idea with new experience.
It should be said also that there exist certain problems with
correspondence between local integrals of motion for
classical and quantum cases in the theory of
asymptotically free models. The reason for that is
very fundamental, that is why our consideration will be
not Hamiltonian one.

Particular (and, probably, personal) interest in this business
is the following. We have very nice formulae for final physical
quantities in both
classical and quantum theory of integrable models.
However existing way of quantization [8](which has clear
Hamiltonian meaning) does not allow to proceed directly from ones
to others, it starts with beautiful things (R-matrix, algebraic
Bethe anzatz),
but then leads through usual jungle with renormalizations and all that.
We would like to have
direct way even if it has less clear meaning. Similar ideas were applied
to study of correlation functions of vertex models of
statistical mechanics in recent works from Kyoto school [9].

Let us consider a particular asymptotically free model.
Namely, let it be the $su(2)$ chiral Gross-Neveu model.
The spectrum of the model
contains one two-component
particle. The two-particle S-matrix is given by [10]:
$$S_{1,2}(\beta_1-\beta_2)=\frac
{\Gamma\bigl(\frac {\beta_1-\beta_2} {2\pi i}\bigr)
 \Gamma\bigl(\frac 1 2 +\frac {\beta_1-\beta_2} {2\pi i}\bigr)}
  {\Gamma\bigl(\frac {\beta_1-\beta_2} {2\pi i}\bigr)
 \Gamma\bigl(\frac 1 2 -\frac {\beta_1-\beta_2} {2\pi i}\bigr)}
\bigl[ \frac {(\beta_1-\beta_2)-\pi i P_{1,2}}
 {\beta_1-\beta_2-\pi i} \bigr], $$
where $P_{1,2}$ is permutation.
If we are interested in local operators which
transform under integer-spin representation
of global $SL(2)$ isotopic group all of them can be
obtained as descendents with respect to different
conservation laws of one nonlocal operator
called ${\frak A} (x_0,x_1)$. The connection with important
local operators will be explained later. The operator
${\frak A}(x_0,x_1)$ is scalar with respect to both
isotopic and Lorenz transformations. In form factor
approach [11] it can be presented as follows:
$$\align
&{\frak A}(x_0,x_1)=\sum_{m,n=0}^{\infty} \frac {1} {m!n!}
\cr &\times \int_{-\infty}^{\infty}d\alpha_1 \cdots
\int_{-\infty}^{\infty}d\alpha_m
\int_{-\infty}^{\infty}d\beta_1 \cdots
\int_{-\infty}^{\infty}d\beta_n\
Z^*_{ \epsilon_ 1}(\alpha_1)\cdots Z^*_{ \epsilon_m}(\alpha_m)\cr
&
\times F(\alpha_m, \cdots, \alpha_1|
\beta_1,\cdots,\beta_n)_{ \delta_1,\cdots, \delta_n}^{ \epsilon_m,\cdots,
\epsilon_ 1}
\ Z^{ \delta_n}(\beta_n)\cdots Z^{ \delta_1}\beta_1)\cr &
\times \text{exp}(\sum i p^{\mu}(\alpha_i)x_{\mu}-
\sum i  p^{\mu}(\beta_i)x_{\mu})
\tag {1.1}\endalign $$
where $Z^*,Z$ are Zamolodchikov-Faddeev
creation-annihilation operators  of particles,
$p_{\mu}(\alpha)$ is one-particle energy-momentum
($p_{\mu}(\alpha)=M(\text{exp}(\alpha)+
(-1)^{\mu}\text{exp}(\alpha))$), $\epsilon,\delta =\pm$
are isotopic indices.
$F$ are form factors. They satisfy the crossing symmetry
requirement:
$$\align & F(\alpha_m, \cdots, \alpha_1|
\beta_1,\cdots,\beta_n)_{ \delta_1,\cdots, \delta_n}^{ \epsilon_m,\cdots,
\epsilon_ 1}
=c_{\delta_1,\delta_1'} \cdots c_{\delta_n,\delta_n'}\\&\times
F(\alpha_m, \cdots, \alpha_1
,\beta_1+\pi i-i0,\cdots,\beta_n+\pi i-i0)^{\epsilon_m,\cdots, \epsilon_ 1,
 \delta_1',\cdots, \delta_n'}
,\ c=i\sigma^2,\cr
 \endalign $$
It should be mentioned that for the
operator ${\frak A}$
the two particle form factor has
a double pole at $ \beta_2=\beta_1+\pi i$ the
corresponding singular part being normalized as
$ (\beta_2-\beta_1 -\pi i)^{-2}c^{\epsilon_ 1, \epsilon_2} $.

The model allows abelian and non-abelian symmetries.
The first are the local conservation laws with
all possible odd spins. The one particle
eigenvalues of the local integrals $I_s$
($s=2m+1,\ m=-\infty,\cdots,\infty$) are equal
to $\text{exp}(s\alpha)$ the spectrum of them
being additive which can be expressed in more formal way
by the following comultiplication formula:
$$ \Delta (I_s)=I_s\otimes 1+1 \otimes I_s $$
The nonlocal symmetries together with $SL(2)$-charges
constitute the Yangian ($Y$) [2]. The generators of $Y$ are
$J_0^a,J_1^a,\ a=1,2,3$ whose action onto one
particle is described by $\sigma^a,\ \alpha\sigma^a$
respectively and the comultiplication is given by
$$\align & \Delta (J_0^a)=J_0^a\otimes 1+1\otimes J_0^a,\cr
  &  \Delta (J_1^a)=J_1^a\otimes 1+1\otimes J_1^a+
  \pi i f^{abc} J_0^b\otimes J_0^c \tag {1.2}\endalign $$
Every operator acting in the space of states
($H_p$) which is the Fock space of particles has descendents
with respect to both local and non-local
symmetries defined through ajoint action.
One can think of the space $H_p$ as of
$$H_{p}=\bigoplus\limits _{n=1}^{\infty} \int
\limits _{\beta _1<\cdots<\beta _{2n}}V_{\beta _1}\otimes\cdots
\otimes V_{\beta _{2n}} \tag {1.3}$$
where $V_{\beta _{i}}$ is the space of the representation $\rho _{\beta _i}$.
The operators $Z^*(\beta),\ Z(\beta)$ either add or remove
one space $V_{\beta}$.
In particular the energy-momentum tensor
and currents are the following descendents of the operator
${\frak A}$:
$$ T_{\mu,\nu} (x_0,x_1)=\epsilon_{\mu,\mu^{\prime}}
                \epsilon_{\nu,\nu^{\prime}}
                P_{\mu^{\prime}} P_{\nu^{\prime}}({\frak A} (x_0,x_1)),\
   j_{\mu}^a (x_0,x_1)=\epsilon_{\mu,\mu^{\prime}} P_{\mu^{\prime}}
J_1^a({\frak A} (x_0,x_1))$$
where $P_0=I_1+I_{-1},\ P_1=I_1-I_{-1}$.
There are more general relations. For example the
density of the local conservation law $I_s$ ($d_s (x_0,x_1)$) is
obtained in the following way:
$$d_s (x_0,x_1)=P_0 I_s ({\frak A} (x_0,x_1))$$
The last relation is very much in common
with the classical relation identifying the
densities of local conservation laws with the
derivatives of logarithm of $\tau$-function.
So, it is reasonable to think of the
operator ${\frak A} (x_0,x_1)$ as of the quantum analog of
$\text{log}(\tau)$. Certainly, this is a formal analogy
and in order to make it more instructive we have to
develop the relation with the classical $\tau$-function.
This is the main goal of the present paper.
For the reasons which will be explained later
the connection in question can hardly be achieved in
Hamiltonian formalism, so, our strategy will be to
find a ``good'' formula in classics in which all the
objects involved in quantum formula (the operator ${\frak A}$,
the form factors, Zamolodchikov-Faddeev operators)
will have their classical analogs. This is what we are going to do in that
paper.
But before proceeding in that direction
we have to remind certain facts about the symmetries of the
quantum model.

\newpage

{\bf 2.The Yangian symmetry.}

In this section the brief account of the recent
developments concerning the dynamical symmetries of
integrable models will be presented.

As it has been already mentioned the model under
consideration allows nonlocal conservation laws $J_1^a$
which together with isotopic charges  $J_0^a$
constitute the infinite dimensional algebra called Yangian [12].
It is explained in [5,6] that for the application to
the description of local and quasilocal operators in the
theory one has to add one more generator ($J_{-1}^a$) to the
algebra. The full algebra generated by  $J_0^a$,  $J_1^a$,
 $J_{-1}^a$ is called the double of Yangian being denoted by
$D(Y)$. One can think of $D(Y)$ as of a deformation of the
affine algebra $\widehat {sl}(2)$ with natural identification of the
generators. This will be explained more precisely later.
The Yangian $Y$ is a subalgebra of $D(Y)$ which acting onto
the fields in the theory creates descendents. The introducing
of the second half of the algebra was originally
motivated by the problem of description of the commutation
relations among these descendents[5].
The commutation relations are described as follows.
Consider a local operator $\varphi (x_0,x_1)$ which can be
for the simplicity taken as invariant under the isotopic algebra.
Now create all possible descendents of the operator acting on it
by arbitrary number of the operators $J_1^a$:
$$\varphi (x_0,x_1)^{a_1,\cdots,a_n}=
J_1^{a_1} \cdots J_1^{a_n}(\varphi (x_0,x_1))$$
We can combine these operators into $\Phi$ considering
$a_1,\cdots,a_n$ as multiindex. So, $\Phi$ is an infinite
column of operators which can be considered as belonging to
$End(H_p)\otimes W$ where $H_p$ is the space of states, $W$ is
the Verma module created by action of the operator  $J_1^a$.
Now, if we consider two towers of operators $\Phi(x_0,x_1)$
and $\Phi(x_0,x_1')$ all the products of their elements
can be combined in the following object:
$\Phi_1(x_0,x_1)\Phi_2(x_0,x_1')$ where two copies of $W$
($W_1$,$W_2$) are considered and $\Phi_1 \in End(H_p)\otimes W_1,\
\Phi_2 \in End(H_p)\otimes W_2$. The commutation relation in question can be
written down in the following way [5]:
$$\Phi_1(x_0,x_1)\Phi_2(x_0,x_1')=
   R_{1,2}\Phi_1(x_0,x_1')\Phi_2(x_0,x_1),\ \text{for}\ x_1>x_1'$$
Here $R_{1,2}$ is $D(Y)$ R-matrix acting in the tensor product
of two Verma moduli. That means that in these relations we effectively
consider the local field $\varphi$ as that annihilated by the
operator $J_{-1}^a$.

It is explained in the papers [6] that the situation can be
inversed. We considered the space of particles and came to the
conclusion that the local and quasilocal operators acting in the space
behave, as far as the commutation relations are considered, as being
combined into Verma moduli. The local operators themselves are identified with
the highest vectors of the Verma moduli.
We can consider now the space of fields instead
of the space of particles. The space of fields ($H_f$) is a suitable
collection of Verma moduli. Particular local field $\varphi_i$ is identified
with a highest vector $|0\rangle _i$ satisfying the requirement
$$J_{-1}^a|0\rangle _i=0$$
We consider also a dual vacuum $\langle 0|$ which is defined by the
relations
$$\langle 0|J_0^a=0,\ \langle 0|J_1^a=0$$
The algebra $D(A)$ allows beside of the infinite
dimensional representations the finite-dimensional ones.
We will be interested in two-dimensional representation ($\rho_{\beta}$)
depending on a parameter $\beta$, the detailed description is
given in [6]. One can define the vertex operators $V^{\epsilon}(\beta)$
which belongs to $End(H_f)\otimes C^2$ ($\epsilon=\pm$ is $C^2$
index). Acting in $H_f$ the operator $V^{\epsilon}(\beta)$ transforms
under the ajoint representation with respect to the
representation $\rho_{\beta}$:
$$ad_x(V^{\epsilon}(\beta))=\rho_{\beta}(x)^{\epsilon}_{\epsilon '}
V^{\epsilon '}(\beta)$$
Then the vacuum expectations of the operators  $V^{\epsilon}(\beta)$
given by
$$ \langle 0|
V^{\epsilon_1}(\beta_1)\cdots V^{\epsilon_{2n}}(\beta_{2n})|0\rangle _i$$
\define\VVV#1{ \langle 0|
V^{\epsilon_1}(\la 1)\cdots V^{\epsilon_{2n}}(\la {#1})|0\rangle }
satisfy the deformed Knizhnik-Zamolodchikov equations, in usual
tensor notation they take a form:
$$ \align & \langle 0|
V(\beta_1)\cdots V(\beta _j+2\pi)\cdots
 V(\beta_{2n})|0\rangle _i=
 \\
&= S_{2n,j} (\beta _{2n}-\beta _{j}-2\pi i)
\cdots S_{j+1,j} (\beta _{j+1}-\beta _{j}-2\pi i)\\
& \times S_{1,j} (\beta _{1}-\beta _{j})
\cdots S_{j-1,j} (\beta _{j-1}-\beta _{j})\\
& \times \langle 0|
V(\beta_1)\cdots V(\beta _j)\cdots
 V(\beta_{2n})|0\rangle _i \tag {2.1}
\endalign $$
As it has been explained in [12] these equations are consistent
with the symmetry property

$$ \align
&S_{i,i+1}(\beta_i - \beta_{i+1}) \langle 0|
V(\beta_{1}),\cdots,V(\beta_i),V(\beta_{i+1}),\cdots,V(\beta_{2n})
|0\rangle _i = \cr &=P_{i,i+1} \langle 0|V(
\beta_{1}),\cdots,V(\beta_{i+1}),V(\beta_{i}),\cdots,
V(\beta_{2n})|0\rangle _i \tag {2.2}
\endalign
$$
Being supplied with the symmetry property (2.2) the equations
(2.1) appear to be the same as the basic requirements for the form
factors. There is an additional equation on residues in the
form factor bootstrap approach [11] which we do not present here.
This equation can be interpreted as a form of operator product
requirement [6].
It is responsible for
the special choice of blocks of the vertex operators.
Namely, the following remarkable identification can be done
$$ F(\beta_1,\cdots,\beta_{2n})^{\epsilon_1,\cdots,\epsilon_{2n}} \sim
\langle 0|V^{\epsilon_1}(\beta_1)\cdots V^{\epsilon_{2n}}(\beta_{2n})|0\rangle
\tag {2.3}$$
where $|0\rangle $ is the highest vector of spin zero, The consequence
of the Verma moduli in the RHS is taken as follows (we indicate the
spins of the highest vectors)$$0\to 1/2 \to 1 \to 1/2\to 1\to\cdots
\to 1/2 \to 0 \tag {2.4}$$ We put the asymptotical equivalence in
(2.3) for the reasons explained in [6], we shall
briefly explain the point later. The form factor can be considered
as a matrix element of the operator ${\frak A}$ in the space of particles.
More generally the relation (2.3) can be rewritten as follows
$$\langle vac|Z^{\epsilon_1}(\beta_1)\cdots
Z^{\epsilon_{2n}}(\beta_{2n})\varphi_i(0)|vac\rangle \sim
\langle 0|V^{\epsilon_1}(\beta_1)\cdots V^{\epsilon_{2n}}(\beta_{2n})
|0\rangle _i$$
We do not expect other local operators then the descendents under
the action of local integrals on ${\frak A}$, and on the currents $j_{\mu}^a$
to exist in the theory. There is also parafermionic field of spin
$\frac 1 4$ with respect to Lorentz transformations (kink field).
{}From the point of view of the representation
theory these three fields correspond to spin 0,1,1/2 highest
vectors respectively. The sequence of the Verma moduli
in form factors of these operators is the same as in (2.4)
the only difference being that the sequence for currents
(kinks) terminates at spin 1 (1/2) representation.
This is a variant of ``rationality'' of the theory. The
reason for the phenomenon to occur supposed to be in the absence of mixing
of different solution of deformed KZ equation by braiding which
is discussed in details in the papers [6].

Now we are going to explain the asymptotical character of the equivalence
in the relations (2.3). This matter was considered in [6,12].
The situation is briefly as follows.
In LHS of (2.3) we have the analytical solutions of
(2.1) while the RHS contains the vacuum expectation of
$D(Y)$-vertex operators. The RHS can be in principle
calculated directly (without use of KZ) using the
definition of the vertex operators. It is quite
clear that the only thing we are able to get in that
way are certain power series in $\beta$'s. But as it
follows from the explicit formulae for the
analytical solutions they are transcendental functions
of $\beta$'s (the explicit formulae are given in [11]);
roughly, they have the properties of $\Gamma$-functions.
The only reasonable connection between these two
types of objects is asymptotical one. One can have in
mind the following analogy: $\Gamma$-function (RHS) and
its asymptotical series (LHS) both satisfy
the functional equation for the  $\Gamma$-function
(dKZ-equation). The difference between the function and
its asymptotics in the case is due to the exponential
in $\beta$ contributions which are considered as the
contribution due to the intermediate state created by local
integrals which should be added to the $D(Y)$ [6].

\newpage

{\bf 3.The classical limit.}

\define \asl {\widehat{sl}(2)}
As it has been already said the Yangian double is
a deformation of the algebra $\asl$. Let us ignore for
the moment the physical content, and consider the formal
aspects of the limit. The typical relation for us will be (1.2).
By rescaling of $J_1^a$ the relation can be rewritten as
$$\align & \Delta (J_0^a)=J_0^a\otimes 1+1\otimes J_0^a,\cr
  &  \Delta (J_1^a)=J_1^a\otimes 1+1\otimes J_1^a+
  \frac h 2 f^{abc} J_0^b\otimes J_0^c  \tag {3.1}\endalign $$
where $h$ is arbitrary constant. Certainly , the
generator $J_{-1}^a$ should be rescaled as well.
Now consider $h$ as Planck constant and take
the limit $h\to +i0$ ( we prefer considering
imaginary $h$, so, probably, temperature is better
analog for it then Planck constant). In the limit the
equations (3.1) turn into trivial comultiplication formulae
for the generators of $\asl$, all other $D(Y)$-relations turn
into $\asl$-ones in  the limit. Also the dKZ equations
turn formally into usual $\asl$ KZ equations on
level zero. It is clear that the rescaling of the
generators we did for the dKZ equations is
equivalent to the rescaling of the rapidities:
$\beta_i=\frac {2\pi i} {h}\lambda_i$ , $h\to +i0$
while $\lambda_i$ are fixed.
So, formally for the functions
$$f(\laaa {2n})\simeq_{h\to +i0}C(h)F(\frac {2\pi i \lambda_1}{h},\cdots,
\frac {2\pi i \lambda_{2n}}{h}) $$
($C(h)$ is a normalization constant)
the equations (2.1) turn in this limit into

$$\bigl( \frac{d}{d\la i}+ \sum_{i\neq j}r_{i,j}
(\la i - \la j)\bigr)f(\laaa {2n})=0 $$
where $r$ is the classical r-matrix:
$$r_{i,j}
(\la i - \la j)\bigr)=\frac {\sigma^a_i\otimes \sigma^a_j}
 {\la i-\la j}$$
The connection between deformed and undeformed KZ
equation is described explicitly in [13].
Let us outline the basic points.

First, the dKZ equation allows the same number of solutions as
undeformed one. The solution corresponding to the form
factors is a special one. Its particular character is explained by
(2.4), we shall also describe it from other point of view soon.

Second, the solutions are in one-to-one correspondence
in the following sense. Consider some solution of dKZ
for $\beta_1<\beta_2<\cdots <\beta_{2n}$ then its asymptotics
$\beta_i=\frac {2\pi i} {h}\lambda_i$ , $h\to +i0$
is described by a solution of KZ. In the paper [13] the
explicit formulae are presented for solution of
dKZ which correspond to all possible solutions of KZ in
that sense.

Third, there is essential difference in the properties
of the solution of dKZ in comparison with those of KZ:
braiding does not mix different solutions of dKZ.
The difference is explained by the asymptotical character
of the correspondence above. Braiding does not commute with
taking the asymptotics.

The third point mentioned makes a real difference between
the deformed and underformed case. In particular,
in the deformed case it makes sense to consider the
form factors of ${\frak A}$ themselves without consideration
other solutions to dKZ equations. But taking the
asymptotics of the the form factors we have to consider not
only the solution of KZ we obtain, but also those
which are connected with it by braiding. We shall call this part
of solutions of KZ the main part.

We do not give here the explicit formulae for the
solutions of dKZ which can be found in [11,13], but we have to
write down explicitly the solutions of KZ. These
formulae concerns the particular case of zero central
extension which causes essential simplification with
respect to the general case. So, let us explain first
how the simplifications appear.
In Varchenko-Shechtman [14] formulae for the
solutions of KZ one has the following structure:
$$\align \prod_{i<j}(\lambda_i-\lambda_j)^{\frac 1 {2(k+2)}}
\int d\tau_1 \cdots d\tau_k &\prod_{m,i}
(\tau_m-\lambda_i)^{\frac 1 {(k+2)}}\prod_{l<m}
(\tau_l-\tau_m)^{\frac 2 {(k+2)}}\\ &\times
R(\tau_1,\cdots;\lambda_1,\cdots)\tag {3.2}\endalign$$
where $\tau$'s are integration variables (corresponding to
screenings), $R$ is a rational function of its variables.
For $k=0$ the exponents in (3.2) turn into 1/4, 1/2, 1 respectively.
The first one is not so essential, the second causes the
hyperelliptic character of the integrals, the third causes the absence
of branching between the integration variables which simplifies
essentially the choice of contours of integration. Also one
can present $\prod_{l<m}
(\tau_l-\tau_m)$ as Vandermond determinant, and rewriting
properly the function $R$ perform the integration over the columns
of the determinant getting a determinant of single
integrals instead of the multi- integral of determinant.
Let us present the result to which these manipulations
should lead (which
actually has not been obtained in that way, but directly
from the classical limit the solutions of dKZ solutions).

\define\eee#1{\epsilon_1, \cdots, \epsilon_{#1}}
Consider the components of $f(\laaa {2n})$:
$f(\laaa {2n})^{\eee {2n}}$. We are looking for the singlet
solutions of the equations which means in particular that
$\sum \epsilon_i=0$. For each particular component
$f(\laaa {2n})^{\eee {2n}}$ the multiindex $\eee {2n}$
induces a partition of $B=\{1,\cdots,2n\}$ into
$T=\{i_k\}_{k=1}^n:\ \epsilon _{i_k}=+$ and
$T'=B\backslash T=\{j_k\}_{k=1}^n:\ \epsilon _{j_k}=-$. Different
solutions will be parametrized by the sets
$\gaaa {n-1}$ which will be specified later. The solutions look
as follows
$$
\align
&f_{\gaaa {n-1}}(\laaa {2n})^{\eee {2n}}=
\prod_{i<j}(\la i-\la j)^{\frac 1 4}\\
&\times\prod\limits _{i\in T,j\in T'}
(\la i - \la j)^{-1}
\text{det}||\int_{\gamma _i}\zeta_j(\tau|T|T')d\tau \ ||
_{g\times g}
\tag {3.3}\endalign
$$
$\zeta_j$ are the following differentials on the hyper-elliptic surface (HES)
$w^2=P(\tau)\equiv \prod (\tau -\la i)$:
$$ \align &\zeta_j(\tau|T|T')
=\frac{Q_j(\tau|T|T')}{\sqrt{P(\tau)}},\\ &Q_j(\tau)=
\bigl\{ \prod\limits _S (\tau -\la l)
\bigl [\frac{d}{d\tau}\frac { \prod _{S'} (\tau -\la l)}
{\tau^{n-j}} \bigr ]_0+
 \prod\limits _{S'} (\tau -\la l)
\bigl [\frac{d}{d\tau}\frac { \prod _S (\tau -\la l)}
{\tau^{n-j}} \bigr ]_0 \bigr \}\tag {3.4}
\endalign
$$
where $[\ ]_0$ means that only the polynomial part
of the expression in brackets is taken.
The differentials $\zeta_j$ are of the second kind:
they have singularities at $\infty^{\pm}$, but their
residues at the infinities are equal to zero. It should be mentioned that
the singular part of the differential $\zeta_j$ is
independent of the partition of $\Lambda$:
$$\zeta_j(\tau|T|T')-
\zeta_j(\tau|T_1|T_1')=
\text{of the first kind} $$
First kind differentials in our case are of the type:
$\sigma _j=\tau^{j-1}/\sqrt{P(\tau)},\ \ 1\le j \le n-1$.
The contours $\gaaa {n-1}$ are arbitrary cycles on the HES ( notice
that its genus $g$ equals $n-1$ ).

It can be shown that the asymptotics of the form factor $F$
corresponds to the following special choice of $\gaaa {n-1}$:
they are taken as canonical $a$-cycles $a_1,\cdots, a_g$ where
the cycle $a_i$ surrounds the cut between $\la {2i-1}$ and
$\la {2i}$, $i=1,\cdots, n-1$ (we suppose $\lambda$'s to be ordered:
$\la 1 <\la 2<\cdots < \la {2n}$).

Now, let us turn to the problem of braiding. The equations (5.3) are
invariant under the permutation $\la i \leftrightarrow \la j$ and
simultaneous permutation of the  associated spaces. Let us denote
the operation of the analytical continuation
$\la i \leftrightarrow \la {i+1}$ and permutation of corresponding
spaces by $B_{i,i+1}$. Then we are supposed to have a formula of
the type:
\define\gaaaa#1{{\gamma}^{\prime}_1,\cdots,{\gamma}^{\prime}_{#1}}
$$ \align
&B_{i,i+1}f_{\gaaa {n-1}}(\laaa {2n})=\\
&=\sum_{\gaaaa {n-1}} C^{\gaaaa {n-1}}_{\gaaa {n-1}}
f_{\gaaaa {n-1}} (\laaa {2n})
\endalign
$$
where $C$ are some constants. There is one
interpretation of the braiding following from the
formula (3.3) : $\la i \leftrightarrow \la {i+1}$
corresponds to certain element of the modular group $Sp(2g,Z)$
of the HES. Clearly under the braiding the determinants (3.3)
transform under $g$-th exterior power of the vector
representation of the group. It can be shown that this
interpretation implies that the solution $f$ corresponding
to $\gaaa {g}=a_1,\cdots,a_g$  mixes by braiding only
with those solutions which correspond to other choices of
half-basis of homologies ($g$ contours on the surface with
zero intersection numbers). This is the main part of solutions
of KZ defined above. There is the same number of elements
in the main part as the number of independent half-bases, i. e.
$  C_{2g+2}^{g+1}- C_{2g+2}^{g}$ of them.
On the other hand it is well known nowadays
that the braiding is described by the finite-dimensional
quantum group $SL(2)_q$  [15], in our case $q=-1$.
The number of solution coincides
with the the multiplicity of the one-dimensional
representation in the tensor product of $2n$ two-
dimensional representations ot $SL(2)_q$ which is
in generic situation $C_{2n}^{n}-C_{2n}^{n-1}$
(the same as above since $g=n-1$). So, it is
natural to suppose that the main part of solutions
for the case $q=-1$ allows continuation to other $q$
while the rest of solutions is special for $q=-1$.
In what follows we shall need only the main part.
There is a beautiful way of rewriting the formulae (3.3)
for the case of the main part of solutions in
terms of $\theta$-functions. We shall consider that in the
next section.

Now let us turn to the problem of physical interpretation
of the classical limit in question. We have to understand
what does the limit $D(Y) \to \asl $ mean. In that limit, in
particular, the antipode-square automorphism of $D(Y)$ turns
into differentiation for $\asl$:
$$ s^2 \to 1+hD \tag {3.5}$$
We do not denote $D$ by familiar notation ($L_{-1}$) to avoid
a confusion: $D$ is not a derivative in the space-time.
There is a strange thing about this limit. The antipode-square
($s^2$) is
identified with the rotation of the space-time in the theory
[6], for two-dimensional representations (particles)
it corresponds to the shift of rapidity by $2\pi i$.
So, the operation, $s^2$ corresponds to, in the
space-time of QFT model is essentially finite being
of topological character, hence the consideration
of its infinitesimal limit (3.5) does not look very
reasonable. However we do will to
consider the limit because from mathematical point of
view the limits of certain objects (e.g. form
factors) do make sense. Thinking more of this
situation one comes upon an idea that in the limit in question the
very notion of the space-time of the QFT model should
be lost. Form factors are the objects referring to one
point in the space-time, so their limits in this
strange situation might make sense because the notion of
one point might remain (it simply appears to be isolated
from the rest of the world). Certainly form factors also
do not survive in full meaning. First, we deal not with a real
limit, but with asymptotical equivalence. Second, the
difference disappears between two-dimensional representation and
dual to it which contained for $D(Y)$ essential crossing shift,
and the two-dimensional representation can not
be considered as those corresponding to particles.
Thus, the only reasonable guess we can make is that
the QFT in the limit splits into a family
of systems with finite degrees of freedom,
the limits of form factors being somehow connected with
these systems. The situation is difficult also
because we do not expect any connection between these
finite-dimensional system and QFT on Hamiltonian level.

Let us consider the situation in opposite direction.
Suppose we have certain family of classical
systems every of which has a finite number of degrees
of freedom. These systems should be unified through
the fact of possessing the symmetry under $\asl$.
Now we perform a quantization of these systems
which essentially leads only to the quantization
of $\asl$ ($\asl \to D(Y)$). It should be mentioned
that there is a jump in quantization of $\asl$: as far as we got a small $h$ in
$D(Y)$ it can be immediately rescaled to finite one.
So, the quantization provides a finite operation
($s^2$). Now we introduce a space-time, and identify this
operation with the rotation of this space-time.
The coordinates are introduced as those
respecting this interpretation. As soon as it is done
we get the notion of particle, and one space-time
point we started with appears to be able to interact with
others through exchange of particles.
This interpretation seems to be reasonable.
In particular it solves the problem of disagreement
between local conservation laws in classical and quantum
integrable asymptotically free models. In a context
close to the present consideration this problem was discussed in [6].

So, now we have to explain what classical systems with finite number
of degrees of freedom and $\asl$-invariance we have in mind. The
answer is implied by the structure of formulae of the solutions of KZ
on level zero. They are connected with HES. So, it is natural to
suppose that the classical systems in question are stationary
finite-gap solutions [16-17] of classical soliton equation with $\asl$
symmetry. These systems will be described in Section 5.  But before
doing that we have to explain the connection between the solutions of
KZ on level zero and Riemann $\theta$-functions.

\newpage

{\bf 4.KZ equations on level zero and Riemann theta-functions.}

In this section we shall derive a formula which express
the solutions of KZ equations on level zero in terms
of Riemann $\theta$-functions. Let us first fix the
notations and introduce necessary definitions.

We consider the HES $\Sigma$ of genus $g$ with $2g+2$ real branching
points ordered as follows: $\la 1 < \cdots <\la {2g+2}$.
Let us put the cuts between the points $\la {2i-1}, \la {2i}$
for $i=1,\cdots,g+1$.
The $a$-cycles on the surface are taken in canonical way:
the cycle $a_i$ surrounds the cut $\la {2i-1}, \la {2i}$,
for $i=1,\cdots,g$. The $b$-cycle $b_i$ ($i=1,\cdots, g$)
starts from one bank of the cut $\la {2i-1}, \la {2i}$, reaches
the cut $\la {2g+1}, \la {2g+2}$ by one sheet, and then returns
th the other bank of $\la {2i-1}, \la {2i}$ by another sheet.
There are $g$ nonsingular differentials on the HES:
$\sigma_j=\tau ^{j-1}/\sqrt{P(\tau)}$. The normalized
first kind differentials $\omega _j$ are linear combinations of
$\sigma$'s satisfying the condition:
$$\int_{a_i}\omega _j =\delta_{i,j}.$$
The martix of periods $\Omega$ is defined as
$$\Omega_{ij}=\int_{b_i}\omega_j.$$
The period matrix is symmetric due to Riemann bilinear identity.

The second kind differentials possess singularities, but their
residues at the singular points vanish.
There is one type of the second kind differentials of
particular importance. The differentials of this
type are obtained by erasing the dependence on one argument
of the two-differentials
$\omega^2 (x,y)$ defined on $\Sigma\times \Sigma$. The differential
$\omega^2 (x,y)$ possesses the only singularity at the
diagonal being normalized as
$$ \omega^2 (x,y)=\frac 1 {(x-y)^2}dxdy, \qquad x\sim y$$
Being considered as a differential in one variable (say $x$)
it satisfy the normalization condition:
$$\int_{a_i}\omega^2 =0,\qquad \forall i$$
Finally, the differential $\omega^2$ is symmetric:
$\omega^2 (x,y)= \omega^2 (y,x)$.

Riemann $\theta$-function is defined
as follows:
$$\theta (z|\Omega)=\sum_{m\in Z^g}\text{exp}
\{\pi i\ m^t\Omega  m+2\pi i z^tm\}$$
where $z \in C^g$. The periodicity property says
$$\theta (z+\lambda'+\Omega\lambda''|\Omega)=
\EXP\{-\pi i\lambda^{\prime\prime\ t}\Omega\lambda''-2\pi i\lambda'' z\}
\theta (z|\Omega), \ \lambda',\lambda''\in Z^g$$
The $\theta$-function with characteristics is defined
by:
$$\teet { }{z|\Omega}=\text{exp}\bigl\{\pi i\eta^{\prime\prime\ t}
\Omega\eta''+
2\pi i(z+\eta')^t\eta''\bigr\}\theta(z+\eta'+\Omega\eta''|\Omega)$$
where $\eta=(\eta ',\eta '')$ is a characteristic: the vectors $\eta ',\eta ''$
belong to $R^g$ ($R$ is the field of rational numbers).
We will be interested in the case when $\eta$ is a half-
period characteristics which means that $\eta ',\eta ''
\ \in 1/2Z^g$.

The definition of $\theta$-function refers to the
particular choice of $a$ and $b$ cycles. However,
the $\theta$-functions defined with respect to
different choices of the homology bases are connected
due to the modular property of $\theta$-function.
Suppose that we have two homology bases connected
via a transformation from $\Gamma _{1,2} \in Sp(2g,Z)$:
$$\pmatrix A' \\ B'\endpmatrix=
\pmatrix d &c\\b &a \endpmatrix
\pmatrix A \\ B\endpmatrix, \qquad \pmatrix d &c\\b &a \endpmatrix \in
Sp(2g,Z),\
 \matrix \text{diag}\  cd^t\\  \text{diag}\  ab^t\endmatrix
\ \equiv 0 (mod 2) $$
``diag'' means the vector composed of the diagonal entries of the matrix.
The transformation law for $\theta$-functions says that [19]:
$$\theta [\xi] (z'|\Omega ')=\gamma(\text{det}(M))^{1/2}
\text{exp}\bigl\{\pi i \sum_{i<j}z_iz_j
\frac {\partial \text{log}(\text{det}M)}{\partial\Omega_{i,j}}\bigr\}
\teet { } {z|\Omega}\tag {4.1}
$$
where
$M=(c\Omega+d),\  z=M^tz' ,\  \Omega'=(a\Omega+b)(c\Omega+d)^{-1}$, the
characteristics changes as follows
$$\pmatrix \xi '\\ \xi ''\endpmatrix=
\pmatrix a &-c\\ -b &d \endpmatrix
\pmatrix \eta'\\ \eta''\endpmatrix$$
$\gamma$ is irrelevant for our goals 8-th root of unity.

Let us denote the set $\{1,2,\cdots,2g+2\}$ counting the
branching points of $\Sigma$ by $B$.
The most significant property of HES is that they allow one-to-one
correspondence between half-periods
and the subsets of $B$ with even number of elements
(mod identification $T\sim B\backslash T$) [19]. The correspondence
is achieved as follows. Consider the subset $T$ ($\#T=2m$)
and divide it into two subsets $T_1,T_2$ such that
$\#T_1=\#T_2=m$. Now let $T_1=\{i_1,\cdots,i_m\},\
T_2=\{j_1,\cdots,j_m\}$ and associate to $T$ the half-period
characteristic $\tilde{\eta}_T$ such that
$$\tilde{\eta}'_T+\Omega\tilde{\eta}''_T=
\sum_k\int\limits_{\la {i_k}}^{\la {j_k}}\omega \tag {4.2}$$
where $\omega$ is considered as vector composed of first kind differentials.
It can be easily shown that the the ambiguity in dividing of $T$
into subsets and enumerating these
subsets changes $\tilde{\eta}_T$ by full period which can be ignored.
It is convenient to measure the characteristics relative to
Riemann characteristic $\delta$ which is defined as
$$\delta=\left\{ \matrix \tilde{\eta}_U,&\ \text{if}\ g\equiv 1 (mod2)\\
\tilde{\eta}_{\{U\backslash1\}},&\ \text{if}\ g\equiv 0 (mod2)\endmatrix
\right.$$
where $U=\{1,3,5,\cdots\}$. The characteristic $\eta _T$ corresponding to
the subset $T$ such that $\#T\equiv g+1 (mod2)$ is defined by
$$\eta_T=\tilde{\eta}_{T\circ U}$$
where $T\circ U=(T\cup U)\backslash (T\cap U)$.

Our nearest goal is to prove the following

{\bf Proposition.}  Consider the solution of KZ on level
zero which corresponds to $\gaaa {g}=a_1,\cdots,a_g$.
The components of this solution are denoted by \linebreak
$f(\laaa  {2g+2})^{\eee {2g+2}}$.
To every multiindex $\eee {2n}$ a partition of $B$ corresponds
such that $B=T\cup T', \qquad i\in T:\epsilon _i=+,
i\in T':\epsilon _i=-$. Denote the elements of $T$ by $i_p,\
p=1,\cdots,g+1$. The set $T$ is associated with $\theta$-
characteristic $\eta _T$. The following relation holds:
$$f(\laaa {2g+2})^{\eee {2g+2}}=C^{-3}\teet T 0 ^4
\DET ||\partial _p\partial _q\text{log}\teet {T}{0}|| _ {g\times g}
 \tag {4.3}$$
where
$C$ is the constant:
$C=\prod _{i<j}(\la i-\la j)^{\frac 1 4}\Delta,$ $\Delta$
is the following important in future determinant:
$$\Delta=\DET ||\int _{a_i} \sigma _j||_{g\times g}$$
The notations are used:
$$\partial _iF(\cdots,0,\cdots)
 \equiv\frac{\partial}{\partial z_i}
F(\cdots,z_i,\cdots)|_{z_i=0}$$

To prove the proposition we have to explain certain facts
concerning $\theta$-functions on HES.

Tomae formulae says that $\teet T 0 =0$ if $\#T\neq g+1$, and
if $\#T=g+1$ then
$${\teet T 0}^2 =\prod _{i<j,i,j \in T} (\la i -\la j)^{\frac 1 2}
\prod _{i<j,i,j \in B\backslash T} (\la i -\la j)^{\frac 1 2}
\Delta \tag {4.4}$$

The following remarkable relation between the differentials
$\omega^2 $  the derivatives of $\theta$-function at $z=0$
($\theta$-constants) takes place. If $x,y,\eta$ are such that
$\teet {} {\int_x^y\omega}=0$ then [20]:
$$\omega^2(x,y)=-\sum _{i,j} \partial _i\partial _j \text{log}\teet {}{0}
\omega _i(x)\omega _j(y) \tag {4.5}$$

To proceed further we need explicit formulae for
certain differentials. The normalized first kind differential $\omega _i$
is given by $g\times g$ determinant:
$$\omega_i(x)=(-1)^i\Delta ^{-1} \DET \pmatrix
\sigma _1(x),&\int_{a_1}\sigma _1&\cdots & \int_{a_{i-1}}\sigma _{1},
 & \int_{a_{i+1}}\sigma _{1},&\cdots & \int_{a_{g}}\sigma _{1}\\
\vdots&\vdots&\vdots&\vdots&\vdots&\vdots&\vdots\\
\sigma _g(x),&\int_{a_1}\sigma _g&\cdots & \int_{a_{i-1}}\sigma _{g},
 & \int_{a_{i+1}}\sigma _{g},&\cdots & \int_{a_{g}}\sigma _{g}
\endpmatrix
$$
Let us consider now the differential $\omega ^2 (x,y)$
as function of $x$
specifying $y$ to one of the branching points: $y=\la i$.
We take $\xi=\sqrt{(\tau-\la i)}$ as
local parameter in vicinity of $\la i$. The differential
$\omega ^2 (x,\la i)$ is second kind differentil with zero
$a$-periods and fixed singularity: $\omega ^2 (x,\la i)\sim \xi ^{-2},
x\sim \la i$. It can be presented as $(g+1)\times(g+1)$ determinant:
$$\omega ^2(x,\la i)=\Delta ^{-1} (P'(\la i))^{\frac 1 2}\DET \pmatrix
\sigma _1(x),&\int_{a_1}\sigma _1&\cdots
&\cdots & \int_{a_{g}}\sigma _{1}\\
\vdots&\vdots&\vdots&\vdots&\vdots\\
\sigma _g(x),&\int_{a_1}\sigma _g&\cdots
&\cdots & \int_{a_{g}}\sigma _{g}\\
\rho _i(x),&\int_{a_1}\rho _i&\cdots
&\cdots & \int_{a_{g}}\rho _i
\endpmatrix $$
where $\rho _i(\tau)=\frac 1 {(\tau-\la i)\sqrt{P(\tau)}},\qquad
P'(\la i)\equiv\prod _j^{\prime}(\la i-\la j)$.
{}From this formula one derive the following
representation for $\omega ^2 (x,\la i)$ when
$x=\la j$ in terms of $g\times g$ determinant:
$$\omega ^2(\la j,\la i)=\Delta ^{-1}
 \bigl(\frac {P'(\la i)}{P'(\la j)}\bigr)^{\frac 1 2}
\DET \pmatrix
\int_{a_1}\sigma _{2,j}&\cdots
&\cdots & \int_{a_{g}}\sigma _{2,g}\\
\vdots&\vdots&\vdots&\vdots\\
\int_{a_1}\sigma _{g,j}&\cdots
&\cdots & \int_{a_{g}}\sigma _{g,j}\\
\int_{a_1}\rho _{i,j}&\cdots
&\cdots & \int_{a_{g}}\rho _{i,j}
\endpmatrix \tag {4.6}$$
where $\sigma _{l,j}=\sigma _{l}-\la j\sigma _{l-1},\qquad
\rho _{i,j}=\rho _i+(\la j-\la i)^{-1}\sigma _1$.
Combining (4.6) and (4.5) one arrives at
$$\sum \partial _p\partial _q\teet T 0 A_u^pA_v^q\la i ^u\la j ^v=
\Delta ^{-1}P'(\la i)
\DET \pmatrix
\int_{a_1}\sigma _{2,j}&\cdots
&\cdots & \int_{a_{g}}\sigma _{2,g}\\
\vdots&\vdots&\vdots&\vdots\\
\int_{a_1}\sigma _{g,j}&\cdots
&\cdots & \int_{a_{g}}\sigma _{g,j}\\
\int_{a_1}\rho _{i,j}&\cdots
&\cdots & \int_{a_{g}}\rho _{i,j}
\endpmatrix \tag {4.7}$$
where $i ,j\in T$, the matrix $A$ connects normalized
first kind differentials with trivial ones:
$$\omega=A\sigma,\qquad A=||\int _{a_i}\sigma _j||^{-1}\tag {4.8}$$
The reason for the equation (4.7) to hold is that adding
$\int _{\la i}^{\la j}\omega $ to $\eta _T$ we get
odd characteristics corresponding to $T\backslash \{i,j\}$, and
$\teet {T\backslash \{i,j\}}{0}=0$.

Let us return to the solution of KZ equations corresponding to
$\gaaa {g}=a_1,\cdots,a_g$. It is given by (3.3):
$$
\align
&f_(\laaa {2g+2})^{\eee {2g+2}}=
\prod_{i<j}(\la i-\la j)^{\frac 1 4}\\
&\prod\limits _{i\in T,j\in T'}
(\la i - \la j)^{-1}
\text{det}||\int_{a_i}\zeta_j(\tau|T|T')d\tau \ ||
_{g\times g} \tag {4.9}
\endalign
$$
where  $T,T'$ are associated to $\eee g$ in a usual way.
Let us take $g$ elements
from $T$ (say $i_p,\ p=1,\cdots,g$) and multiply the
determinant from (4.9) by Vandermond composed of corresponding $\la {}$'s
the result being
$$ \DET ||F_{p,q}||_{g\times g},\qquad
F_{p,q}=\int _{a_p}\sum _{m=1}^g \zeta _m {\la {i_q}}^{g-m}$$
The following two differentials
are equivalent (differ by a total derivative):
$$\sum _{m=1}^g \zeta _m {\la {i_q}}^{g-m}\sim \mu _q\equiv
  \prod_ {j\in B\backslash T}(\la {i_q}-\la j)
\frac {\prod_{i\in T,i\neq i_q}(\tau-\la i)}{(\tau-\la {i_q})\sqrt{P(\tau)}}
$$
Hence the determinant in (4.9) can be replaced by
$$\frac {1} {\prod\limits _{i<i', i,i'\in T\backslash i_{g+1}}
(\la i -\la {i'})}
\DET (C), \qquad C_{pq}= \int _{a_p}\mu _q \tag {4.10}$$
Let us find the matrix $X$ which satisfies the equation:
$$AX=C \tag {4.11}$$
with $A$ and $C$ given by (4.8),(4.10). Kramer's rule tells that
$$ X_{p,q}=(-1)^p \Delta ^{-1}
\DET \pmatrix
\int_{a_1}\sigma _1&\cdots&\cdots
&\cdots & \int_{a_{g}}\sigma _1\\
\vdots&\vdots&\vdots&\vdots&\vdots\\
\int_{a_1}\sigma _{p-1}&\cdots &\cdots
&\cdots & \int_{a_g}\sigma _{p-1}\\
\int_{a_1}\sigma _{p+1}&\cdots &\cdots
&\cdots & \int_{a_g}\sigma _{p+1}\\
\vdots&\vdots&\vdots&\vdots&\vdots\\
\int_{a_1}\sigma _{g}&\cdots &\cdots
&\cdots & \int_{a_g}\sigma _{g}\\
\int_{a_1}\mu_q &\cdots &\cdots
&\cdots & \int_{a_g}\mu _q
\endpmatrix $$
Let us now consider the determinant of the
matrix $X$. Multiplying it by Vandermond composed of
$\la {i_k},\ k=1,\cdots,g$ one gets
$$\DET(X)=\frac {1} {\prod\limits _{i<i', i,i'\in T\backslash i_{g+1}}
(\la i -\la {i'})}\DET (\tilde{X})$$
where
$$\tilde {X}_{p,q}=
 \Delta ^{-1}
\DET \pmatrix
\int_{a_1}\sigma _{2,p}&\cdots&\cdots
&\cdots & \int_{a_{g}}\sigma _{2,p}\\
\vdots&\vdots&\vdots&\vdots&\vdots\\
\int_{a_1}\sigma _{g,p}&\cdots &\cdots
&\cdots & \int_{a_g}\sigma _{g,p}\\
\int_{a_1}\mu_q &\cdots &\cdots
&\cdots & \int_{a_g}\mu _q
\endpmatrix  \tag {4.12}$$
It is easy to show that $\mu _q$ in (4.12) can be replaced
by $P'(\la {i_q})\ \rho _{i_p,i_q}$. Combining that
with (4.7) and taking into account that $\DET(A)=\Delta ^{-1}$
one gets:
$$\DET (X)=\prod\limits _{i<i', i,i'\in T\backslash i_{g+1}}
(\la i -\la {i'})
\DET ||\partial _p\partial _q\text{log}\teet {T}{0}||_{g\times g}$$
Calculating determinants of RHS and LHS of formulae (4.11)
and having in mind (4.8) we obtain:
$$\DET||\int _{a_i}\zeta _j||=\Delta
\DET ||\partial _p\partial _q\text{log}\teet {T}{0}|| \tag {4.13}$$
Together with Tomae formulae it provides the representation
for the solutions of KZ on level 0:
$$f(\laaa {2g+2})^{\eee {2g+2}}=C^{-3}\teet T 0 ^4
\DET ||\partial _p\partial _q\text{log}\teet {T}{0}||_{g\times g}$$
where $T$ is related to $\eee {2g+2}$ as explained above,
$C=\prod _{i<j}(\la i-\la j)^{\frac 1 4}\Delta$.
That proves the proposition above.

The formula (4.3) has very beautiful meaning. It relates the
solution of KZ equations which are differential equation with respect
to moduli of HES (the positions of branching points) in terms of
derivatives on the Jacobian. It would be nice to prove directly
that (4.3) satisfies KZ using the heat equation for the
$\theta$-function.

\define\fff#1#2{f_{#1}(\laaa {#2})^{\eee{#2}}}
Now suppose that we took a solution corresponding to
other choice of half basis: $\gaaa {g}=a'_1,\cdots,a'_g$.
Denote this solution by $f_{A'}(\laaa {2g+2})$. Clearly,
all the reasonings above are applicable to $f_{A'}$
the difference being that the final formula has to
contain $\theta$-functions defined with respect to
the half-basis $A'$. In order to rewrite the answer in terms of
canonical $\theta$-functions (those corresponding to $A$)
we have to use the formula (4.1). After simple computations one gets:
$$\align &f_{A'}(\laaa {2g+2})^{\eee {2g+2}}=\\ &\DET (M)
C^{-3}\teet T 0 ^4
\DET ||\partial _p\partial _q\text{log}\teet {T}{0}
+\pi i\frac {\partial
\text{log}(\text{det}M)}{\partial\Omega_{i,j}}||\endalign$$
where
$$\pmatrix A' \\ B'\endpmatrix=
\pmatrix d &c\\b &a \endpmatrix
\pmatrix A \\ B\endpmatrix, \qquad \pmatrix d &c\\b &a\endpmatrix \in \Gamma
_{1,2}. $$
The matrix $M$ is defined as
$M=(c\Omega+d)$.

In Section 6 we shall need another representation for the
solutions of KZ in terms of $\theta$-functions. The author
was unable to prove this representation completely, so we shall
formulate it as conjecture and present reasonings in favour of
it.

{\bf Conjecture 1.} For every half-basis $A'$ a polynomial $Q_{A'}$
in $\partial _i$ of total degree $2g$ exists such that
$$ f_{A'}(\laaa {2g+2})^{\eee {2g+2}}=C^{-3}{\teet T 0}^2
Q_{A'}{\teet T 0}^2  \tag {4.14}$$
where the connection between $\eee {2g+2}$ and
$T$ ia as usual, the coefficients of $Q_{A'}$ might be
complicated, but they do not depend on $T$.

Let us explain why we assume the representation to exist.
To this end we have to understand what kind of
constants depending on $T$ is a candidate for being
presentable in the form ${\teet T 0}^2
Q{\teet T 0}^2$ for some polynomial in $\partial _i$
and $T$ corresponding to even non-singular characteristics
($\#T=g+1$).
Riemann $\theta$-functions on HES satisfy many relations
which follow from the combination of Riemann relations with
the peculiar properties of HES. In particular, the following
relations hold for $\teet {} z $ ($\eta$ is even, nonsingular,
i.e. such that $\teet {} 0 \neq 0$)[19]:
$$\sum\limits _{S\subset B,\#S=g+1, 1\in S}(-1)^{S\cap (U\circ T)}
{\teet S z}^2 {\teet S 0}^2=0 \tag {4.15}$$
where $T$ is arbitrary subset of $B$ satisfying the
requirements:$\#T\equiv g+1(mod\ 2)$, $\#T\neq g+1,\ 1\in T$.
We put the requirement $1\in S$  in order
to avoid the summation over $B\backslash S$.
So, if the set of constants is presentable in the form
$Q\teet S 0 ^2$ then they should satisfy this system of
relations being placed instead of $\teet S z ^2$.
The opposite should be also true:
if a set of constants enumerated by $S$ satisfies the
system (4.15) then they can be presented in the form
$Q\teet S 0 ^2$. In elliptic case ($g=1$) there
is only one set $T$ satisfying the requirements above:
$T=B$. For this set the relation (4.15) turns into
$${\theta _{0,0}(0)}^2{\theta _{0,0}(z)}^2=
{\theta _{1,0}(0)}^2{\theta _{1,0}(z)}^2+
{\theta _{0,1}(0)}^2{\theta _{0,1}(z)}^2$$
in usual notation for the elliptic $\theta$-functions
with characteristics [19]:
$$\theta _{1,0}(z)=\teet {\{1,2\}} z,\
\theta _{0,0}(z)=\teet {\{1,3\}} z,\
\theta _{0,1}(z)=\teet {\{1,4\}} z.$$

Let us prove that the solutions of KZ for arbitrary
choice of $A'$ satisfy the relations (4.15). First, let us
check that  ${\teet S 0}^2$ given by Tomae formulae
satisfy the requirements. Substitute ${\teet S 0}^2$ given by (4.4)
into (4.15) instead of both ${\teet S 0}^2$ and
${\teet S z}^2$ and divide the relation by
$\Delta ^2 \prod _{i<j}(\la i-\la j)$ the result being
$$\sum\limits _{S\subset B,\#S=g+1, 1\in S}(-1)^{S\cap (U\circ T)}
\frac {1} {\prod\limits _{i\in S,j\in B\backslash S}
(\la i -\la j)}=0 \tag {4.16}$$
These identities are proven by consideration of
residues at $\la i=\la j,\ \forall i,j$.

Now let us present the solution of KZ (4.9) as multidimensional
integral:
$$
\align
&f_{A'}(\laaa {2g+2})^{\eee {2g+2}}=
\prod_{i<j}(\la i-\la j)^{\frac 1 4}\\
&\times\prod_{i\in S, j\in B\backslash S}
(\la i - \la j)^{-1}
\int_{a'_1}d\tau _1\frac 1 {\sqrt{P(\tau _1)}}\cdots \int_{a'_g}d\tau _g
\frac 1 {\sqrt{P(\tau _g)}}\DET ||Q_i(\tau _j|S|S')||_{g\times g}
\endalign
$$
where $Q_i$ are given by (3.4), $S$ is related to $\eee {2g+2}$ as usual.
Let us denote $\DET ||Q_i(\tau _j)||_{g\times g}$
by $X_g(\taaa {g}|S|S')$ indicating the dependence on genus $g$ and
subsets $S,S'$ (recall that $S'=B\backslash S$).
We shall prove that the solutions of KZ
satisfy (4.15) being substituted as ${\teet S 0}^2{\teet S z}^2$
if we prove that
$\prod_{i\in S, j\in S'}
(\la i - \la j)^{-1}X_g(\taaa {g}|S|S')$
satisfy them. So, we have to prove that
$$\sum\limits _{S\subset B,\#S=g+1, 1\in S}(-1)^{S\cap (U\circ T)}
\prod_{i\in S, j\in S'}
(\la i - \la j)^{-1}X_g(\taaa {g}|S|S')=0 \tag {4.17}$$
In fact the proof of the identity does not differ much from the
proof of (4.16). Again it is sufficient to check the
cancellation of the residues at $\la i=\la j$ because
$X_g(\taaa {g}|S|S')$ is of degree $g$ with respect to
any $\la {}$. Suppose $l\in S,\ m\in S'$ then the polynomial
$X_g(\taaa {g}|S|S')$ satisfies the following recurrent
relation (classical version of the relations from [11]):
$$\align &X_g(\taaa {g}|S|S')\big | _{\la l=\la m}=
\prod\limits _{k=1}^g (\tau _k-\la l)
\sum\limits _{k=1}^g (-1)^k
\bigl\{\frac {\partial}{\partial \tau _k}
\prod\limits _{p\in B\backslash \{l,m\}}(\tau _k-\la p)\bigr\}\\&\times
X_{g-1}(\taaa {k-1},\tau _{k+1},\cdots,\tau _g|S\backslash l|S'\backslash m)
\endalign $$
The point is that the coefficients in the relation are
independent of $S$. That provides the possibility of inductive
proof of (4.17).

It would be nice to have explicit formulae for the
polynomials $Q$ in (4.14). The author has not succeeded to
get them. Presumably the formulae can be found from (4.3)
using Fay identities [20].

To finish this section let us present explicit formulae for the
elliptic case. The canonic way for construction of homology basis
is as follows: the cycle $a$ surrounds the cut
between $\la 1,\ \la 2$, the cycle $b$ starts from upper
bank of this cut, reaches the upper bank of the cut
$\la 3,\ \la 4$ by one sheet, then moves to another sheet
and starting with the lower bank of $\la 3,\ \la 4$
returns to the lower bank $\la 1,\ \la 2$.
There are two independent choices of $a$-cycle:
$a=a$ or $a'=b$ which are connected by the following
element from $Sp(2,Z)$:
$$\pmatrix 0 & 1\\-1 & 0 \endpmatrix  $$
So, there are two independent solutions to KZ:
$$\align & f_a(\laaa 4)^{\eee 4}=C^{-3}\teet T 0 ^4
\partial ^2 \text{log}\teet T 0, \\
& f_{a'}(\laaa 4)^{\eee 4}=C^{-3}\teet T 0 ^4 K \bigl[
\partial ^2 \text{log}\teet T 0 +K^{-1}\bigr]  \tag {4.18}\endalign$$
where $K$ is full elliptic integral:$K=\frac 1 {\pi i}
\int\limits _{\la 1}^{\la 2}\frac {d\tau}
{\sqrt{P(\tau)}}$,the correspondence between $\eee 4$,  $T$ and conventional
notations for elliptic $\theta$-functions are as follows:
$$\align &\{++--\},\{--++\}\to\{1,2\} \to \{1,0\},\\
&\{+--+\},\{-++-\}\to\{1,4\} \to \{0,1\},\\
&\{+-+-\},\{-+-+\}\to\{1,3\} \to \{0,0\} \tag {4.19} \endalign $$
The solution can be also rewritten as (4.14) with
$$Q_a=\partial ^2,\qquad Q_{a'}=K\partial ^2+1$$
since $\partial\teet {} 0=0$ for even $\eta$.

\newpage

{\bf 5.Pragmatic view of finite gap integration.}

\define\MM {\frak M}

Consider an integrable equation with infinitely many
times $t_1,t_2,\cdots$. To any time $t_i$ the $M$-operator
$\MM _i$ is attached. The $M$-operators satisfy zero curvature
condition:
$$\frac {\partial}{\partial t_i}\MM_j-\frac {\partial}{\partial t_j}
\MM_i=\bigl[ \MM_i,\MM_j \bigr] \tag {5.1}$$
We consider $\widehat{sl}(2)$-invariant case (one can think of
nonlinear Shr\"odinger model for example), so $\MM_i$ is
traceless $2\times 2$
matrix depending on spectral parameter $\lambda$ such that
$$\text{tr}\MM_i(\lambda)=0, \ \MM_i(\lambda)=\sum_{k=0}^i
\lambda^k m_{i,k},$$
also we normalize it by the requirement $m_{i,i}=\sigma^3$.

The coefficients of the $M$-operators are dynamical variables.
Finite-gap integration [16-18] deals with the situation of stationary
solutions which means that there is a time $t_n$ on which
the dynamical variables do not depend:
$$\frac {\partial}{\partial t_n} \MM_i=0, \ \ \forall i$$
That means that the $M$-operator $\MM_n$ satisfies the equation:
$$\frac {\partial}{\partial t_i} \MM_n =\bigl[\MM_i,\MM_n], \ \ \forall i $$
hence the determinant of $\MM_n$ is an integral of motion with
respect to all the times. Certainly this determinant can be
presented as follows:
$$ \text{det}\MM_n(\lambda )=\prod _{k=1}^{2n} (\lambda-\lambda_k)$$
That suggests that the problem in question is closely related
to the problem of parametrization of all matrices $M(\lambda )$
(we omit the index $n$) which satisfy the following
requirements: $M(\lambda)$ is traceless $2\times 2$ matrix depending
on $\lambda$ as polynomial of degree $n$ with fixed
senior coefficient (equal $\sigma^3$) and given determinant
($ \text{det}M(\lambda )=\prod _{k=1}^{2n} (\lambda-\lambda_k)$).
This problem was considered in the last century (Jacobi, Riemann), the
solution can be found for example in Mumford's book [19]. Let
us describe it in the terms appropriate for our further goals.

The matrix in question is degenerate at the points $\lambda=
\lambda_j$, also it is traceless, hence $M(\lambda_j)$ is
a Jordan cell:
$$M(\lambda_j)=\psi_j \otimes \bar{\psi}_j\qquad
\bar{\psi}_j\equiv \psi_j^t \sigma^2$$
\define\bpsi#1{\bar{\psi}_{#1}}
for some vector $\psi_j$. Let us parametrize $M(\lambda)$
by the set $\psi_j,\ j=1,\cdots , 2n$ (these vectors are
not independent as we shall see soon). Construct the interpolation
$$M^{\prime}(\lambda)=\sum_{k=1}^{2n} \frac { \prod_{p\neq k}
 (\lambda -\lambda_p)} { \prod_{p\neq k}
 (\lambda_k -\lambda_p)}M_k$$
Definitely, $M^{\prime}$ is degenerated at the given points
with given values, but its degree in $\lambda$ equals $2n-1$
instead of $n$ required. So, we have to kill $n-1$ senior
coefficients of $M^{\prime}$, also we have to take into
account that $\lambda ^n$ is supposed to enter with given
coefficient $\sigma ^3$. After some simple manipulations
these requirements lead to the following system of quadratic
relations for the components $a_i , b_i$
of the vectors $\psi _i$:
$$ \align
& \sum _{k \in T} \frac {1}{\prod_{j \in T} ^{\prime}
(\lambda_k - \lambda _j)}a_k^2=0,
\cr
&
  \sum _{k \in T} \frac {1}{\prod_{j \in T} ^{\prime}
(\lambda_k - \lambda _j)}b_k^2=0,\cr
&  \sum _{k \in T} \frac {1}{\prod_{j \in T} ^{\prime}
(\lambda_k - \lambda _j)}a_k b_k=1  \tag {5.2}
\endalign $$
for any subset $T$ of $n+1$ elements of the set $\{1,2,\cdots,2n\}$.
This system of equations leaves only $n-1$ independent parameters
(which could be, certainly, calculated from the very beginning).
The important point is that due to Riemann identities and
Tomae formulae the relations allow a parametrization
in Riemann $\theta$ functions on the hyperelliptic surface
$\tau ^2=P(\lambda) \equiv \prod _{k=1}^{2n} (\lambda-\lambda_k)$.
The solution to the system (5.2) looks as follows:
$$a_j=C \frac{\theta [\eta _j] (r+z)}{\theta[ \delta] (z)
\theta[\eta _j] (r)},\
b_j=C \frac{\theta [\eta _j] (r-z)}{\theta[ \delta] (z)
\theta [\eta _j] (r)} \tag {5.3} $$
where $\eta _j$ is the theta-characteristic corresponding
to the branching point $\lambda_j$: $\eta _j=\tilde {\eta}_{1,j}$ with
$\tilde {\eta} _T$ defined by (4.2). The variable $ z=(z_1,\cdots, z_g)$ is
the parameter on the Jacobian (genus $g=n-1$),
$$r=g\int _ {\lambda_1}^{\infty ^+} \omega, $$
$\infty ^+ $ is one of two infinities on the surface,
$ \omega $ is the vector composed of first kind differentials,
$\delta$ is Riemann constant, finally, $C$ is the same as in (4.3).

We would like to make two comments on the above formulae
in order to clarify them from two points of view.
First point is the connection with more familiar
in the context of finite-gap integration object, namely
with Baker-Akhiezer function. The BA function is an
eigenvector of $M(\lambda)$:
$$M(\lambda)\psi (\lambda)=m(\lambda) \psi(\lambda)$$
where $m$ is corresponding eigenvalue. Certainly, for
generic $\lambda$ there are two solution to this equation
which means that the BA function is defined on the surface $\Sigma$.
But for $\lambda = \lambda _i$ the matrix $M(\lambda)$
is degenerate, it has only one eigenvector corresponding to
zero eigenvalue. This eigenvector is exactly our $\psi _i$.
Thus,
$$\psi _i=\psi (\lambda _i)$$
The BA-function $\psi (x)$ ($x=\lambda ^{\pm}$
is a point on the surface) is written in terms of
$\theta$-functions through $\theta (\int ^x \omega + \cdots)$,
[16-18] when $x$ coincides with one of the branching points this
expression turns into $\theta$-function with corresponding
half-period characteristics because the integral of $\omega$
taken between two branching points on hyperelliptic surface is a
half-period.

The second point is the direct connection with well
known addition theorems for $\theta$-functions in the
elliptic case ($n=2$, there are four branching points).
In that case the following identification can be done with
usual $\theta$-functions [19]:
$$
\align &\theta [\eta _1] (z)=\theta_{0,0}(z),\
\theta [\eta _2] (z)=\theta_{0,1}(z),
\cr &\theta [\eta _3] (z)=\theta_{1,1}(z),\
\theta [\eta _4] (z)=\theta_{1,0}(z)
\endalign $$
Also the following simple variant of Tomae formulae
holds:
$$ \align &\theta_{0,0}(0)^2=[(\lambda_1-\lambda _3)
(\lambda_2-\lambda _4)]^{\frac 1 2}K,\
\theta_{0,1}(0)^2=[(\lambda_1-\lambda _4)
(\lambda_2-\lambda _3)]^{\frac 1 2}K \cr
&\theta_{1,0}(0)^2=[(\lambda_1-\lambda _2)
(\lambda_3-\lambda _4)]^{\frac 1 2}K,\
\theta_{1,1}(0)^2=0
\endalign $$
where $K$ is the full elliptic integral
$K=\frac 1 {\pi i}\int _a d\tau /\sqrt{P(\tau)}$, $a$-cycle surrounds
the cut between the points $\lambda_1$ and $\lambda_2$.
The ratios of $\theta$-functions at $r$ are easy
to calculate:
$$ \theta [\eta _j] (r)^2=\text{Const}\
P^{\prime}(\lambda _j)^{-\frac 1 2},\
P^{\prime}(\lambda _j)
\equiv \prod _{k \ne j}(\lambda _j-\lambda _k)$$
Taking all that into account one realizes that for the
parametrization (5.3) considered the
equations (5.2) turn, for example for $T=\{1,2,4\}$ into
$$ \align
& \theta _{0,0}(0)^2 \theta _{0,0}(r+z)^2-
 \theta _{0,1}(0)^2 \theta _{0,1}(r+z)^2-
 \theta _{1,0}(0)^2 \theta _{1,0}(r+z)^2=0, \cr
& \theta _{0,0}(0)^2 \theta _{0,0}(r-z)^2-
 \theta _{0,1}(0)^2 \theta _{0,1}(r-z)^2-
 \theta _{1,0}(0)^2 \theta _{1,0}(r-z)^2=0, \cr
& \theta _{0,0}(0)^2 \theta _{0,0}(r+z) \theta _{0,0}(r-z)-
 \theta _{0,1}(0)^2 \theta _{0,1}(r+z)\theta _{0,1}(r-z)-\cr
& \theta _{1,0}(0)^2 \theta _{1,0}(r+z)\theta _{1,0}(r-z)=
    \theta _{1,1}(z)^2 \theta _{1,1}(r)^2
\endalign $$
The equations (5.2) for other subsets $T$ produce other
known identities
for $\theta$-functions. In the case of generic $n$ the situation is
similar: in parametrization (5.3) the equations (5.2) turn into
certain special cases of the Riemann identities (Frobenius formulae [19]).

Returning to the integrable models we conclude the
following: the formulae (5.3) provide the parametrization
of $M(\lambda)$ on the torus ( Jacobian), that
is
why the times $t_i$ should be related to $z_j$ as
$$t_i=\sum_{j=1}^g c_i^jz_j$$
the constants $c_i^j$ refer to the particular integrable
equation which is not essential for us.

Let us consider now the same situation from different
point of view. What will follow is an extraction from
Adler-Reyman-Semenov-Tian-Shansky approach[21,22].
\define\Aggg{\frak g}
Consider the loop algebra $\Aggg=\widehat{sl}(2)$ with generators
$J_m^a,\ a=1,2,3;\ m$ is an integer. The generators satisfy the
relations:
$$ [J_m^a,J_n^b]=f^{a,b}_c J_{n+m}^c$$
The algebra contains two subalgebras: $\Aggg _-$ generated by
$J_m^a$ with $m<0$ and $\Aggg_+$ generated by $J_m^a$ with
$m \geq 0$. Evidently, the algebra allows finite-dimensional
representation $\rho _ {\lambda}$:
$$\rho _{\lambda}(J_m^a)=\lambda ^m \sigma ^a $$
By $G=\widehat{SL}(2)$ we denote a groop whose Lie algebra
coincides with $\Aggg$. This group contains two subgroups:
$$G_+=\text {exp} (\Aggg _+),\ G_-=\text {exp} (\Aggg _-)$$

Consider now the matrix $M(\lambda)$ of the same type as
above (traceless, polynomial in $\lambda$ of degree $n$
with fixed senior coefficient) and introduce the action of
$G$ on  $M(\lambda)$ as follows:
$$g(M(\lambda))=\rho _ {\lambda}(g)M(\lambda)
 \rho _ {\lambda}(g)^{-1},\ \text{for}\ g \in G \tag {5.4}  $$
The subgroups $G_+,G_-$ acting on $M(\lambda)$ generate
the orbits $O_+,O_-$ of the form:
$$ \align
& g_+(M(\lambda))=\sum_{k=0}^{\infty} m'_k \lambda ^k,
\cr &  g_-(M(\lambda))=\sum_{k=-\infty}^n m''_k \lambda ^k,\
m''_n=\sigma _3 \endalign $$
The algebraic interpretation of the integrable models is
explained as the problem of description of the intersection
$J$ of the orbits $O_+,O_-$. Clearly this
intersection is composed of the matrix of the same type
as $M(\lambda)$ itself, so the connection with the matters
discussed above is manifest. How to describe the
intersection in question in algebraic way? If $M'\in J$
then there exist $g_+,g_-$ such that
$$\align &g_+(M(\lambda))=\rho _ {\lambda}(g_+)M(\lambda)
 \rho _ {\lambda}(g_+)^{-1}= \cr
&=g_-(M(\lambda))=\rho _ {\lambda}(g_-)M(\lambda)
 \rho _ {\lambda}(g_-)^{-1}  \tag {5.5}\endalign $$
It is evident from these equations that the
matrix $g(\lambda)=  \rho _ {\lambda}(g_-)^{-1}\rho _ {\lambda}(g_+)$
commutes with $M(\lambda)$. The only possibility for the
matrix to commute with $M(\lambda)$ which is supposed to be
not degenerate for generic $\lambda$ is to be a function of $M(\lambda)$.
Fortunately, additional dependence on $\lambda$ makes
the situation nontrivial. We can consider $g(\lambda)$
of the form:
$$g(\lambda)=\text {exp}\{\sum t_iM_i(\lambda)\} \tag {5.6} $$
where $M_i(\lambda)=\lambda ^{-i}M(\lambda)$.

Now let us inverse the reasonings which means to start
with $g(\lambda)$ in the form (5.6), and to try to
construct $J$. Evidently to this end we need to
solve the following Riemann problem: present
$$ g(\lambda)=\text {exp}\{\sum t_iM_i(\lambda)\}=
 \rho _ {\lambda}(g_-)^{-1} \rho _ {\lambda}(g_+)  \tag {5.7}$$
where the Loran series for $ \rho _ {\lambda}(g_+)$,
($ \rho _ {\lambda}(g_-)$) contain only
positive (negative) powers. It is clear that only
those $M_i$ are essential for which $0<i<n$, others
can be directly moved to either $g_+$ or $g_-$, and do
not contribute to (5.5). So, we deal with the dependence
on $n-1$ times which can be, actually, identified
with the parameters on the Jacobian considered above.

Suppose the problem (5.7) is solved. Then we can introduce
``times'' dependent matrices $M(\lambda, t_1,\cdots,t_g)$
as being dressed according to (5.5) by $g_+$ or
$g_-$ from (5.7). It easy to show that the projections
of $\lambda^{-i}M(\lambda, t_1,\cdots,t_g)$ onto $\Aggg _+$
denoted by $\MM _i$ satisfy the equations:
$$\frac {\partial}{\partial t_i}\MM_j-\frac {\partial}{\partial t_j}
\MM_i=\bigl[ \MM_i,\MM_j \bigr]$$
which coincide with (5.1).

We would like to emphasize the
importance of the vectors $\psi_i$ from the algebraic point
of view. The BA function itself does not transform in a
reasonable way under the transformations (5.4), but its
values at the branching points ($\psi _i$) do transform
under finite-dimensional representation of $G$ when
$M$ transforms under (5.4):
$$ \psi _i \to G(\lambda_i)\psi_i$$
This nice property of $\psi_i$ will be important in what follows.

To finish this section let us write down explicitly
the equations in terms of $m_i^a$:
$$\frac {\partial}{\partial t_k}m_j^a=
\sum\limits _{q=\text{max}(0,k+j-n)}^jf^{abc}m_{j-q+k}^bm_q^c \tag {5.8}$$

\newpage

{\bf 6.Tau-function and KZ equations.}

The most mysterious object in the theory of
classical integrable equations is that of $\tau$-function [23-26].
Hirota observed that the integrable classical equations can be
rewritten as follows. Let us consider certain function $\tau$,
depending upon the arguments $t=\{t_1,t_2,\cdots\}$
and introduce the notations:
$$D_{i_1}\cdots D_{i_k}\tau\cdot\tau (t)\equiv\frac{\partial}{\partial x_{i_1}}
\cdots\frac{\partial}{\partial x_{i_k}}
\tau(t+x)\tau(t-x)\bigl|_{x=0} \tag {6.1}$$
Then according to Hirota the integrable equations can be
written as
$$P(D)\tau\cdot\tau=0 \tag {6.2}$$
for certain ``polynomial'' in $D$. We shall be interested mostly not in
$\tau$-function itself, but in the linear space of functions $\Tau$
which is defined as a space spanned by the functions
$[D_{i_1}\cdots D_{i_k}\tau\cdot\tau (t)]/\tau (t)^2
$. Hirota equations (6.1)
imply certain linear dependence in the space.

The main achievement of Kyoto school in that direction
is the formula expressing $\tau$-function in terms of
vacuum expectation in the space of highest weight representation for
the algebra $\asl$ on level $1$ [24]. Consider the Cartan subalgebra
generated by $I_k\equiv J_k^3, \ k>0$. Then the formula takes place
$$\tau (t)=\langle \Omega|g\text{exp}(\sum _k t_kI_k)|\Omega\rangle$$
where  $|\Omega\rangle$ is the highest vector of the representation
on level $1$, $\langle \Omega|$ is the dual vector, $g$ is the
element of central extended group $\widehat{SL}(2)$ which
specifies a particular solution of the equation.
We want to relate the KZ equations on level zero to
$\tau$-function. The formula (6.2) is both good and bad for our goals.
Good thing is that (6.2) links the integrable equations to the
highest weight representations which are involved in KZ. Bad thing
is that the formula (6.2) on the one hand and our KZ
on another deal with different central extensions of $\asl$.
We shall take the good thing as a hint of possible
connection between $k=0$ KZ and $\tau$-function, and
forget about the bad thing.
It should be said also that for the case we are
interested in (that of finite-gap integration) the description
of $g$ is rather complicated and indirect.

What is $\tau$-function in the finite-gap case? The answer is well known:
it is essentially the $\theta$-function $\tede z,\ t_i=\sum _kc_i^kz_k$, where
$c$ are some coefficients. How is it connected with BA-function?
The usual answer is that it coincides with the
value of BA-function at certain point ($\infty $).
This answer does not
satisfy us because we do not like to use BA-function at generic
point for its unclear algebraic properties, we want to deal
with the values of BA-function at the branching points ($\psi _j$)
only.

The space $\Tau$  is finite-dimensional in
finite-gap case. Let us consider the example $g=1$. All the times
$t_i$ are proportional to one variable on the Jacobian $z$.
The $\tau$-function is $\theta _{1,1}(z)$. The space $\Tau$
is generated by $1$ and $\frac
{D^2\tau\cdot\tau}{\tau ^2}=\frac{\partial ^2}{\partial z ^2}
\text{log}\theta _{1,1}(z)$.
The Hirota equation reads as
$$D^4\tau\cdot\tau+c_1D^2\tau\cdot\tau+c_2\tau ^2=0$$
which turns into well-known differential equation for
$\theta$-function (equivalent to the
equation for the Weierestrass $\frak{P}$-function):
$$\frac{\partial ^4}{\partial z ^4}
\text{log}\theta _{1,1}(z) +6(\frac{\partial ^2}{\partial z ^2}
\text{log}\theta _{1,1}(z))^2+c_1\frac{\partial ^2}{\partial z^2}
\text{log}\theta _{1,1}(z)+ c_2=0$$
for some $c_1,c_2$.

Our further strategy will be in putting together
the space $\Tau$, the vectors $\psi _i$ and the
solutions of $k=0$ KZ equations the idea being the following:
$\psi _i$ transforms under finite-dimensional representation of $\asl$,
$\tau$ is connected with highest weight representation, hence
they should be put together via KZ equations.
To start let us
consider the following object:
$$v(z)\equiv\bpsi {1,\epsilon _1}(z)\cdots \bpsi {2n,\epsilon _{2n}}(z)
\VVV {2n}$$
it should be said that $v(z)$ is not a function but rather a set of functions
because different
choices of blocks of vertex operators
(intermediate Verma moduli) are possible. This
is the same as different choices of
cycles $A'$ for $\fff {A'}{2n}$:
$$\VVV {2n} _{some \quad  block}=
\fff {some  A'} {2n}$$
So, if necessary we shall indicate a particular component of $v$
as $v_{A'}$.
The first important property of the functions $v(z)$ is that
their dependence on $z$ is governed by ``free'' dynamics
in Verma module.

Let us take the matrix $M(\la {})$ as described in the
previous section:
$$M(\la {})=\sum\limits _{i=0}^n \la {}^i m_i^a\sigma ^a,\
m_n^a=\delta _{a,3},\
 \DET (M(\la {}))=\prod _{i=1}^{2n}(\la {}-\la j) $$
\define\MH{\widehat{M}}
Associate to the matrix $M(\la {})$ the element of $\frak{g}$
$$\MH =\sum\limits _{i=0}^n m_i^a J_i^a$$
such that $M(\la{})=\rho _{\la{}}(\MH)$. Similarly, the algebra elements
$\MH _j$ can be introduced:
$$ \MH _j=\sum\limits _{i=0}^n m_i^a J_{i-j}^a, \qquad \rho _{\lambda}(\MH _j)
=M_j(\la {})$$
Clearly, $\MH _j$ commute among themselves:
$$[\MH _i ,\MH _j]=0$$
Let us consider $M(\la {})$ as a starting point for the
procedure of the previous section, i.e. $M(\la {})$ is
the stationary $M$-operator corresponding to the moment $t_i=0$.
To the matrix $M(\la {})$ we attach the set of vectors
$\psi _i,\ i=1,\cdots,2n$ as it is explained above. The following
important statement holds.

{\bf Proposition} The $z$-dependence of $v(z)$ is governed
by ``free'' dynamics in the Verma module:
$$\align &
v(z)=\bpsi {1,\epsilon _1}(z)\cdots \bpsi {2n,\epsilon _{2n}}(z)
\VVV {2n}=
\\ & \bpsi {1,\epsilon _1}(0)\cdots \bpsi {2n,\epsilon _{2n}}(0)
{ \langle 0|
V^{\epsilon_1}(\la 1)\cdots V^{\epsilon_{2n}}(\la {2n})
\EXP (\sum _{i=1}^{n-1}\MH _i t_i)|0\rangle }\tag {6.3}\endalign $$
where $z$ and $t$ are connected by some linear transformation:
$t_i=c_i^j z_j$.

In what follows we shall allow ourselves
inaccuracy using both notations: $\psi _i(z)$
and $\psi _i(t)$ assuming that they coincide when
$z$ and $t$ are properly related.
As it was explained in the previous section the functions
$\bpsi i$ satisfy the following equation:
$$\frac{\partial}{\partial t_k}\bpsi i (t)=
\bpsi i(t)\MM (\la i,t)$$
where the $t$-dependence of $\MM$ is due to (5.8).
This equation allows to express all the higher derivatives
of $\bpsi {}$ at the moment $t=0$
in terms of derivatives of $\MM$.
Let us prove the equations (6.3) comparing all the derivatives
at $t=0$ of LHS and RHS.

With first derivative it is trivial, in LHS one has:
$$\align & \frac {\partial}
{\partial t_i}  \bigl (\bpsi {1,\epsilon _1}(t)\cdots
 \bpsi {2n,\epsilon _{2n}}(t) \langle 0|
V^{\epsilon_1}(\la 1)\cdots V^{\epsilon_{2n}}(\la {2n})|0\rangle
  \bigr)|_{t=0}=\\ &=\sum\limits _{p=1}^{2n}
 \bpsi {1,\epsilon _1}(0)\cdots
 \bpsi {p-1,\epsilon _{p-1}}(0)
 \bpsi {p,\epsilon _p'}(0)
 \bpsi {p+1,\epsilon _{p+1}}(0)
 \cdots
 \bpsi {2n,\epsilon _{2n}}(0)\\ &\times
\MM _{i,\epsilon _p}^{\epsilon _p'}(\la p) \langle 0|
V^{\epsilon_1}(\la 1)\cdots V^{\epsilon_{2n}}(\la {2n})|0\rangle
\endalign $$
In RHS one has:
$$\align
&\frac{\partial}{\partial t_i}
\bpsi {1,\epsilon _1}(0)\cdots \bpsi {2n,\epsilon _{2n}}(0)
\langle 0|
V^{\epsilon_1}(\la 1)\cdots V^{\epsilon_{2n}}(\la {2n})
\EXP (\sum _{i=1}^{n-1}\MH _i t_i)|0\rangle |_{t=0}=\\
&=\bpsi {1,\epsilon _1}(0)\cdots \bpsi {2n,\epsilon _{2n}}(0)
 \langle 0|
V^{\epsilon_1}(\la 1)\cdots V^{\epsilon_{2n}}(\la {2n})
\MH _{i,+}|0\rangle =\\ &=
\sum\limits _{p=1}^{2n}
 \bpsi {1,\epsilon _1}(0)\cdots
 \bpsi {p-1,\epsilon _{p-1}}(0)
 \bpsi {p,\epsilon _p'}(0)
 \bpsi {p+1,\epsilon _{p+1}}(0)
 \cdots
 \bpsi {2n,\epsilon _{2n}}(0) \\ &\times
\MM _{i,\epsilon _p}^{\epsilon _p'}(\la p) \langle 0|
V^{\epsilon_1}(\la 1)\cdots V^{\epsilon _{2n}}(\la {2n})|0\rangle
\endalign $$
Evaluating the RHS we used the notation $\MH_{i,+}$ for the
projection of $\MH_i$ onto $\frak{g}_+$; recall that $J_m^a|0\rangle=0,\
 m< 0$. We moved $\MH _{i,+}$ to the left using the properties
of the vertex operators, $\MH _{i,+}$ annihilates the left vacuum, also
$\MM _i(\la p)=\rho _{\la p}(\MH _{i,+})$.
Less trivial computation is that for the second derivatives.
In LHS one gets
$$\align &\frac {\partial}
{\partial t_i}\frac {\partial}
{\partial t_j}
\bigl (\bpsi {1,\epsilon _1}(t)\cdots
 \bpsi {2n,\epsilon _{2n}}(t) \langle 0|
V^{\epsilon_1}(\la 1)\cdots V^{\epsilon_{2n}}(\la {2n})|0\rangle
  \bigr)|_{t=0}=\\ &=\sum\limits _{p< q}^{2n}
 \bpsi {1,\epsilon _1}(0)\cdots
 \bpsi {p-1,\epsilon _{p-1}}(0)
 \bpsi {p,\epsilon _p'}(0)
 \bpsi {p+1,\epsilon _{p+1}}(0)
 \cdots \\&\times
\bpsi {q-1,\epsilon _{q-1}}(0)
 \bpsi {q,\epsilon _q'}(0)
 \bpsi {q+1,\epsilon _{q+1}}(0)
 \bpsi {2n,\epsilon _{2n}}(0)
 \\ & \times\bigl\{\MM _{i,\epsilon _p''}^{\epsilon _p'}(\la p)
\MM _{j,\epsilon _q}^{\epsilon _q''}(\la q)+
\MM _{i,\epsilon _q''}^{\epsilon _q'}(\la q)
\MM _{j,\epsilon _p}^{\epsilon _p''}(\la p)\bigr\}
 \langle 0|
V^{\epsilon_1}(\la 1)\cdots V^{\epsilon_{2n}}(\la {2n})|0\rangle+\\
&+\sum\limits _{p=1}^{2n}
 \bpsi {1,\epsilon _1}(0)\cdots
 \bpsi {p-1,\epsilon _{p-1}}(0)
 \bpsi {p,\epsilon _p'}(0)
 \bpsi {p+1,\epsilon _{p+1}}(0)
 \cdots
 \bpsi {2n,\epsilon _{2n}}(0)\\&
\times\bigl\{\MM _{i,\epsilon'' _p}^{\epsilon _p'}(\la p)
\MM _{j,\epsilon _p}^{\epsilon _p''}(\la p)+
\partial _i
\MM  _{j,\epsilon _p}^{\epsilon _p'}(\la p) \bigr\}\langle 0|
V^{\epsilon_1}(\la 1)\cdots V^{\epsilon_{2n}}(\la {2n})|0\rangle
\endalign $$
To transform the RHS one has to deal with $\MH _i\MH _j|0\rangle$
which can be evaluated due to the equations (5.8):
$$\align &
\MH _i\MH _j|0\rangle=(\MH _{i,+}+\MH _{i,-})\MH _j|0\rangle=\\ &
=\MH _{i,+}\MH _{j,+}|0\rangle+
\partial _i\MH _{j,+}|0\rangle \tag {6.4}\endalign$$
Now moving $\MH _{i,+}\MH _{j,+}$ and
$\partial _i\MH _{j,+}$ to the left one gets (6.4).
The consideration of general case is quite similar to the
case of the second derivative which is the most demonstrative one.

Thus, we have shown that the functions $v(z)$ possess the
nice property (6.3) which makes them similar to $\tau$-function.

It should be mentioned that the starting $M(\la {})$ can be
taken in especially simple way. For example we can divide the set of
$\la i$ into two subsets ($B=S\cup S', \ \#S=\#S'=n$) and define
$M$ as follows:
$$M(\la {})=\pmatrix
\prod\limits _{i\in S}(\la{}-\la i) +
\prod\limits _{i\in S'}(\la{}-\la i), &
\prod\limits _{i\in S}(\la{}-\la i) -
\prod\limits _{i\in S'}(\la{}-\la i)\\
\prod\limits _{i\in S'}(\la{}-\la i) -
\prod\limits _{i\in S}(\la{}-\la i), &
-\prod\limits _{i\in S}(\la{}-\la i) -
\prod\limits _{i\in S'}(\la{}-\la i)\endpmatrix $$
In that case the vectors $\psi _i$ are also simple.

We have to connect the functions $v(z)$ with some familiar objects.
There should be a clever way to get the result which will be
announced soon. But we do not know that way which makes us to
proceed to calculations. Let us consider the elliptic case.
Recall that the covectors $\bpsi i$ are given in the elliptic case by:
$$\align &\bpsi 1 (z) \simeq\bigl(\frac{\theta _{0,0}(r+z)}{\theta _{1,1}(z)},\
\frac{\theta _{0,0}(r-z)}{\theta _{1,1}(z)}\bigr)\\ &
\bpsi 2 (z) \simeq\bigl(\frac{\theta _{0,1}(r+z)}{\theta _{1,1}(z)},\
\frac{\theta _{0,1}(r-z)}{\theta _{1,1}(z)}\bigr),\\&
\bpsi 3 (z) \simeq\bigl(\frac{\theta _{1,1}(r+z)}{\theta _{1,1}(z)},\
\frac{\theta _{1,1}(r-z)}{\theta _{1,1}(z)}\bigr)\\ &
\bpsi 4 (z) \simeq\bigl(\frac{\theta _{1,0}(r+z)}{\theta _{1,1}(z)},\
\frac{\theta _{1,0}(r-z)}{\theta _{1,1}(z)}\bigr)\endalign $$
In what follows we shall perform calculations up to constants which
will be outlined by using $\simeq$. There are two solutions
to KZ equations
in elliptic case which are given by are given by
$$\align & f_a(\laaa 4)^{\eee 4}\simeq\teet T 0 ^2
\partial ^2 \teet T 0 ^2, \\
& f_{a'}(\laaa 4)^{\eee 4}\simeq\teet T 0 ^2  \bigl[
K\partial ^2  +1\bigr]\teet T 0 ^2 \endalign$$
the correspondence between $\eee 4$,  $T$ and conventional
notations for elliptic $\theta$-functions is given by (4.19).
This formulae suggest to generalize the expression for $v(z)$
substituting in it $\teet T 0^2\teet T x ^2$ instead of the
solutions of KZ. Doing that one gets the following:
$$\align &v(z,x)\simeq\theta _{1,1}(z)^{-4}\\ &\times
\bigl[\theta _{1,0}(0)^2\theta _{1,0}(x)^2
\{\theta _{0,0}(r+z)\theta _{0,1}(r+z)\theta _{1,1}(r-z)\theta _{1,0}(r-z)+\\
&+\theta _{0,0}(r-z)\theta _{0,1}(r-z)\theta _{1,1}(r+z)\theta
_{1,0}(r+z)\}+\\&+
\theta _{0,1}(0)^2\theta _{0,1}(x)^2
\{\theta _{0,0}(r+z)\theta _{0,1}(r-z)\theta _{1,1}(r-z)\theta _{1,0}(r+z)+\\&+
\theta _{0,0}(r-z)\theta _{0,1}(r+z)\theta _{1,1}(r+z)\theta _{1,0}(r-z)\}+\\&+
\theta _{0,0}(0)^2\theta _{0,0}(x)^2
\{\theta _{0,0}(r+z)\theta _{0,1}(r-z)\theta _{1,1}(r+z)\theta _{1,0}(r-z)+\\&+
\theta _{0,0}(r-z)\theta _{0,1}(r+z)\theta _{1,1}(r-z)\theta
_{1,0}(r+z)\}\bigr]
\endalign $$
This expression can be simplified via Riemann identities the result
being quite simple:
$$v(z,x)\simeq\frac {\theta _{1,1}(z+x)\theta _{1,1}(z-x)}
{\theta _{1,1}(z)^2}$$
which leads to the following nice formulae for $v_a(z)$,$v_{a'}(z)$:
$$v_a(z)\simeq\frac{D^2\tau\cdot\tau}{\tau ^2}(z),\qquad
v_{a'}(z)\simeq K\frac{D^2\tau\cdot\tau}{\tau ^2}(z)+1$$
where $\theta _{1,1}$ is denoted by $\tau$, $D$ is Hirota derivative.

Thus by convoluting $\bpsi i$ and the solutions of KZ in the elliptic
case we got
exactly the basis of the space $\Tau$.
Combining that with the proposition above
we also realize that the dynamics in the space $\Tau$ is
governed by the free dynamics in the Verma module. We would suppose
that this nice connection holds in generic case as well.
In order to prove that we have to use the Conjecture 1 from Section 4
together with the following

{\bf Conjecture 2.} The following identity (up to neglectable
constants) for
the $\theta$-functions on HES holds:
$$\align  \sum\limits _{T\subset B,\#T=g+1}\teet T 0^2&
\teet T x^2 \prod\limits _{j=1}^{2g+2}\teet j {r
+\epsilon _j z}\simeq\\&\simeq\tede z^{2g}\tede {z+x}\tede {z-x}
\endalign $$
where $T$ and $\{\epsilon _i\}$ are connected in usual way.

The author strongly believes in this conjecture, still
he has been unable to prove it.

If the conjectures 1,2 are both true in generic case
then the following relation should take place:
$$\Tau\ni\frac {Q_{A'}\tau\cdot\tau}{\tau^2}=
\bpsi {1,\epsilon_1}\cdots\bpsi {2n,\epsilon_{2n}}
f_{A'}(\laaa {2n})^{\eee {2n}} \tag {6.5}$$
where $\tau (z)=\tede z$.
We would suppose also that there are enough $\theta$-
constants hidden in $f_{A'}$, i.e. that all the
independent Hirota derivatives of $\tau$ can be obtained
by linear combinations of (6.5) for different $A'$.

The formula (6.5) establishes the relation between $\bpsi i$
which transform under finite-dimensional representations
of $\asl$ and the space of Hirota derivatives of $\tau$-function
(the space of classical fields) via the solutions of
KZ equations on level 0 which we can call classical
form factors. This is exactly the formula we were looking for,
its similarity with quantum formula (1.1) is manifest.

To finish this section we would like to present one more
interesting formula:
$$\align &
\langle 0|
V^{\epsilon_1}(\la 1)\cdots V^{\epsilon_{2n}}(\la {2n})
\MH^{2n}\EXP(\sum t_i \MH _i)|0\rangle =\\&=
\psi _ {1}^{\epsilon _1}(0)\cdots \psi _ {2n}^{\epsilon _{2n}}(0)
\bigl\{\bpsi {1,\epsilon '_1}(t)\cdots \bpsi {2n,\epsilon '_{2n}}(t)
\langle 0|
V^{\epsilon '_1}(\la 1)\cdots V^{\epsilon '_{2n}}(\la {2n})
|0\rangle |_{t=0}\bigr\}
=\\&=\psi _ {1}^{\epsilon _1}(0)\cdots \psi _ {2n}^{\epsilon _{2n}}(0)
v(t)\endalign $$
This formula is due to the fact that
$$\rho_{\la i}(\MH)^2=(\psi _i\otimes \bpsi i)^2=0$$
Notice that as usual $\MH$ and $\laaa {2n}$ must be connected:
$\DET(\rho_{\la i}(\MH))=$\linebreak $\prod (\la {}-\la i)$.

\newpage

{\bf 7.Conclusions.}

Let us return to the very beginning of the paper.
We were wondering whether a formula similar to
(1.1) exists in classics. We have answered this question
in the last section of the paper. Really, the
formula (6.5) is very much similar to (1.1). It connects
classical local fields (Hirota derivatives of $\tau$-function)
with the tensor product of finite-dimensional representations
of tha affine algebra via the classical form
factors (the solutions of KZ on level 0).
So, the similarity of the formulae (1.1) and (6.5)
is described by:
$$\align \psi _i,\ \bpsi i &\leftrightarrow\ Z(\beta _i),\ Z^*(\beta _i)\\
\Tau&\leftrightarrow\ {\frak A}\\
f\ &\leftrightarrow\ F \endalign $$
The quantum formula needs also the summation over all
particle states which can be interpreted as summation over all
HES, the number of particles being the related to genus
of surface, the rapidity being considered as the positions
of branching points, i.e. the moduli of the surfaces. Notice that
in the classical limit the difference between the finite-
dimensional representation and conjugated one (which were
responsible for particles and antiparticles in the quantum case)
essentially disappears.

So, we would suppose the following procedure of quantization to
be possible. We start with the family of finite-gap solutions
of integrable model with $\asl$ symmetry. Different solutions
are parametrized by the intersections of the orbits of $\frak{g}_{\pm}$
as it is explained in the Section 5. These intersections are parametrized
by the matrix $M(\la {})$, but we prefer to parametrize them by
the set $\{\psi _i\}$ connected with $M(\la {})$ as it is
explained above. The vectors $\psi _i$ transform under
finite-dimensional representation of the affine algebra.
We can take
$$H_{cl}=\bigoplus\limits _{n=1}^{\infty} \int
\limits _{\la 1<\cdots<\la {2n}}V_{\la 1}\otimes\cdots
\otimes V_{\la {2n}} \tag {7.1}$$
($V_{\la {}}$ is the space of the representation $\rho _{\la {}}$)
as a completion of the manifold of classical finite-gap solutions.
Not every vector from this space is good for the classical
solutions but only those which can be presented in the form
$\psi _1\otimes\cdots\otimes\psi _{2n}$
with the set $\{\psi _i\}$
associated to a matrix $M(\la {})$ as described above.
For every particular element of this form
the classical fields are given by convolution with classical
form factor.
In fact the space (7.1) coincides with the space of states
of quantum model $H_p$ (1.3). Also the quantum form factor is
a quantization of the classical one which is essentially due to
the quantization of the
algebra underlying the theory ($\asl\to D(Y)$).
So, we propose that the following way for quantization should be possible:
we extend the set of classical solutions to the space $H_{cl}$
and identify it with the space of states of quantum model, the rest
of quantization is in quantizing of symmetry algebra.
It should
be strongly emphasized, however, that the space-time of the
quantum model has nothing to do with the classical ``times'' $t_i$,
it appears as a result of quantization as it is explained in
Section 3.

Let us finish with several remarks.
The finite-gap solution constitute in the classical theory
nice but small subset of solutions. Generally, there are
infinite-gap solutions in which the finite-gap ones are of
measure zero. The infinite-gap solutions are rather ugly ones,
no reasonable theory is available for them. So, since for the
quantization only finite-gap solutions are needed, we have a nice
example of usual phenomena: quantization takes everything good
from the classical theory and forgets about bad things typical for it.
There is a puzzling
connection between the matters discussed in the present
paper and those considered in [27] in the connection with the
theory of strings (this remark is due to H.Oogury).
The constructions of the present paper can be generalized to
$su(N)$ case for which the algebra $\widehat{sl}(N)$ and
the Riemann surfaces with the branching points of $N$-th
order are responsible, the form factors of corresponding
$su(N)$-chiral Gross-Neveu are given in [11]. But probably
this is not the best way for the generalization. We would better
proceed to the consideration of arbitrary Riemann surfaces
which are associated to KP-equation. This case should cover
all the $su(N)$-invariant Gross-Neveu models and,
probably, will lead to something essentially new.

{\bf Acknolegement.} The work was partly done under
Institute of Physics Fellowship at I.Newton Institute, Cambridge.
The discussions with I.Cherednik, T.Eguchi, L.Faddeev, A.Kirillov,
T.Miwa, H.Oogury, M.A. Semenov-Tian-Shansky were helpful.

\newpage

\def\no#1#2\par{\item{#1.}#2\par}
\def\jr#1{{\nineit#1}}
\def\book#1{{\nineit#1}}
\def\vl#1{{\ninebf#1}}

{\bf References}

\no 1
M. L\"uscher,
\jr{Nucl. Phys. B}
\vl{135}
(1978), 1

\no {2}
D. Bernard, \jr{CMP}
\vl{137}
(1991), 191

\no 3
D. Bernard, A. LeClair,
\jr{CMP}
\vl{142}
(1991), 99

\no 4
G. Felder, A. LeClair,
\jr{IJMPA}
\vl{7}, suppl.1
(1992),

\no 5
A. LeClair, F.A. Smirnov,
\jr{IJMPA}
\vl{7}
(1992),2997

\no 6
F.A. Smirnov,
\jr{IJMPA}
\vl{7},
suppl.2
(1992)

\no 7
I.B. Frenkel, N.Yu. Reshetikhin,
\jr{CMP}
\vl{146}
(1992), 1

\no 8
L.D. Faddeev, E.K. Sklyanin, L.A. Takhtajan,
\jr{Teor.Math.Phys.}
\vl{40}
(1980), 688

\no 9
M. Jimbo, K. Miki, T. Miwa, A. Nakanishiki,
\book{Correlation functions of XXZ model for $\Delta <-1$}
RIMS preprint 887 (1992)

\no {10}
A.B. Zamolodchikov. Al.B. Zamolodchikov,
\jr{Annals of Physics}
\vl{120}
(1979), 253

\no {11}
F.A. Smirnov,
\book{Form Factors in Completely Integrable Models of Quantum
Field Theory}, Adv. Series in Math. Phys. 14,
World Scientific (1992)

\no {12}
V.G. Drinfeld,
\jr{Sov. Math. Dokl.}
\vl{32}
(1985) 254.

V.G. Drinfeld,
Quantum Groups,
\jr{Proceedings of the International Congress}

\jr{of Mathematicians,}
Berkeley, CA, (1987) 798

\no {13}
F.A. Smirnov,
\book{Remarks on deformed and undeformed KZ equations},
RIMS preprint-860 (1992)

\no {14}
V.V. Schechtman, A.N. Varchenko,
\book{Integral Representations of $N$-point
Conformal Correlations in the
$WZW$ Model},
Max-Plank-Institute preprint
(1987)

\no {15}
A. Tsuchiya, Y. Kanie,
\book{In: Conformal Field Theory and
Solvable Lattice Models,
Adv. Stud. Pure. Math.}
\vl{16}
(1988) 297.

\no {16}
A.R. Its, V.B. Matveev,
\jr{Teor.Math.Phys.}
\vl{23}
(1975), 51

\no {17}
B.A.Dubrovin, V.B. Matveev, S.P. Novikov,
\jr{Russian Math. Surveys}
\vl{31},
(1976) 59

\no {18}
I.M. Krichever,
\jr{Russian Math. Surveys}
\vl{32},
(1977), 185

\no {19}
D. Mumford,
\book{Tata Lectures on Theta}
\vl{1,2},
Birkh\"auser
(1983)

\no {20}
J.D. Fay,
\book{Riemann functions on Riemann surfaces},
Lecture Notes in Mathematics
\vl{352}
(1973)

\no {21}
M.A. Semenov-Tian-Shansky,
\jr{Publ.RIMS Kyoto Univ.}
\vl{21}
(1985), 1237

\no {22}
L.D. Faddeev, L.A. Takhtajan,
\book{ Hamiltonian methods in the theory of solitons},
Springer,
(1987)

\no {23}
R. Hirota,
\book{in: Lecture Notes in Mathematics},
\vl{515}, Springer (1976)

\no {24}
E. Date, M. Jimbo, M. Kashiwara, T. Miwa,
\book{in: Nonlinear Integrable Systems},
World Scientific (1983)

\no {25}
G. Segal, G. Wilson,
\jr{Publ. I.H.E.S.}
\vl{61}
(1985), 1

\no {26}
I. Cherednik,
\jr{Funk. Anal. i Prilozh.}
\vl{17}
(1983), 93

\no{27}
V.G. Knizhnik,
\jr{CMP}
\vl{112}
(1987), 567

\enddocument